\pdfoutput=1
\documentclass[twocolumn]{aastex63}
\usepackage{longtable}
\usepackage{tablefootnote}
\usepackage{graphicx}
\usepackage{multirow}
\usepackage{here}
\received{}
\revised{\today}
\accepted{}
\submitjournal{ApJ}

\shorttitle{Kinematics of XZ Tau}
\shortauthors{Ichikawa et al.}

\begin{document}
\title{Misaligned Circumstellar Disks and Orbital Motion of the Young Binary XZ Tau}

\correspondingauthor{Takanori Ichikawa}
\email{k9145506@kadai.jp}

\author{Takanori Ichikawa}
\affiliation{Department of Physics and Astronomy, Graduate School of Science and Engineering, Kagoshima University, 1-21-35 Korimoto, Kagoshima,Kagoshima 890-0065, Japan}

\author{Miyu Kido}
\affiliation{Department of Physics and Astronomy, Graduate School of Science and Engineering, Kagoshima University, 1-21-35 Korimoto, Kagoshima,Kagoshima 890-0065, Japan}

\author{Daisuke Takaishi}
\affiliation{Department of Physics and Astronomy, Graduate School of Science and Engineering, Kagoshima University, 1-21-35 Korimoto, Kagoshima,Kagoshima 890-0065, Japan}

\author[0000-0001-9368-3143]{Yoshito Shimajiri}
\affiliation{Department of Physics and Astronomy, Graduate School of Science and Engineering, Kagoshima University, 1-21-35 Korimoto, Kagoshima,Kagoshima 890-0065, Japan}
\affiliation{National Astronomical Observatory of Japan, Osawa 2-21-1, Mitaka, Tokyo 181-8588, Japan}

\author{Yusuke Tsukamoto}
\affiliation{Department of Physics and Astronomy, Graduate School of Science and Engineering, Kagoshima University, 1-21-35 Korimoto, Kagoshima,Kagoshima 890-0065, Japan}


\author[0000-0003-0845-128X]{Shigehisa Takakuwa}
\affiliation{Department of Physics and Astronomy, Graduate School of Science and Engineering, Kagoshima University, 1-21-35 Korimoto, Kagoshima,Kagoshima 890-0065, Japan}
\affiliation{Academia Sinica Institute of Astronomy and Astrophysics, 11F of Astro-Math Bldg, 1, Sec. 4, Roosevelt Rd, Taipei 10617, Taiwan}



\begin{abstract}
We report our analyses of the multi-epoch (2015--2017) ALMA archival data of the Class II binary system XZ Tau at Bands 3, 4 and 6.
The millimeter dust continuum images show compact, unresolved ($r \lesssim$ 15 au) circumstellar disks (CSDs) around the individual binary stars; XZ Tau A and B, with a projected separation of $\sim$ 39 au.
The $^{12}$CO (2-1) emission associated with those CSDs traces the Keplerian rotations, whose rotational axes are misaligned with each other (P.A. $\sim$ -5$\degr$ for XZ Tau A and $\sim$ 130$\degr$ for XZ Tau B).
The similar systemic velocities of the two CSDs ($V_{\rm LSR}$ $\sim$ 6.0 km s$^{-1}$) suggest that the orbital plane of the binary stars is close to the plane of the sky.
From the multi-epoch ALMA data, we have also identified the relative orbital motion of the binary.
Along with the previous NIR data, we found that the elliptical orbit
($e$ = 0.742$^{+0.025}_{-0.034}$, $a$ = 0$\farcs$172$^{+0\farcs002}_{-0\farcs003}$, and $\omega$ = -54.2$^{+2.0}_{-4.7}\degr$) is preferable to the circular orbit.
Our results suggest that the two CSDs and the orbital plane of the XZ Tau system are all misaligned with each other, and possible mechanisms to produce such a configuration are discussed.
Our analyses of the multi-epoch ALMA archival data demonstrate the feasibility of time-domain science with ALMA.
\end{abstract}


\section{Introduction} \label{sec:intro}
Most ($\gtrsim$50$\%$) solar-type stars belong to binaries \citep{Raghavan10}.
Young (Class I, II) binary systems consist of a pair of circumstellar disks (hereafter CSDs) surrounding individual stars \citep{Maury10,Chen13,Lim16a,Lim16b}.
Several Class I and II binaries are also found to have circumbinary disks (CBDs) surrounding the entire binary systems \citep{Tang14,Dutrey14,Tobin16,Takakuwa14,takakuwa17,takakuwa20}.
Classically, close ($\lesssim$100 au) binaries are considered to be formed via fragmentation of a common disk, precursor of the CBD \citep{Kratter08,Machida08}. 
In such a case, the CSDs should be aligned on the same plane \citep[$e.g.$][]{Lim16a}.
Recent ALMA observations of young binaries have found, however, that the two CSDs can be significantly ($\gtrsim$60$\degr$) misaligned with each other \citep{Jensen14,Brinch16,Kurtovic18}.
Such misalignments could imply that the simple disk fragmentation scenario is not sufficient to fully explain binary formation.
Recent theoretical studies have indeed shown that the disks can be misaligned by the effect of turbulence \citep[$e.g.$][]{Bate10,Offner10,Matsumoto15}.
The turbulence causes local angular momenta to form disks around protostars, by which the CSDs can be misaligned \citep{Takaishi20}.
XZ Tau is a close binary system composed of two T-Tauri stars (XZ Tau A and B), located in the L1551 region at a distance of 146 pc \citep{Roccatagliata20}.
The L1551 region contains Class I protostellar binaries of L1551 IRS 5 and NE \citep{Takakuwa14,takakuwa17,takakuwa20}, and a planet-forming Class I-II source HL Tau \citep{ALMA15,Yen17,Wu18,Yen19a,Yen19b}.
The projected separation between XZ Tau A and B is $\sim$0$\farcs$3 ($\sim$39 au) \citep{Haas90,Torres09}.
The 7-mm continuum observations of XZ Tau with the Very Large Array (VLA) by \cite{Carrasco09} detected the tertiary companion XZ Tau C at a projected distance of 0$\farcs$09 ($\sim$13 au) from XZ Tau A, but the subsequent follow-up observations posed a question on the existence of the tertiary \citep{Forgan14,ALMA15}.
Multi-epoch Hubble Space Telescope (HST) observations revealed the bubble-like outflow driven from XZ Tau \citep{Krist97,Krist08}.
ALMA observations of XZ Tau in the $^{12}$CO (1--0) emission have confirmed the presence of the molecular counterpart of the outflow, and the blue- and redshifted emission are located to the southwest and the northeast at P.A. $\sim$20$\degr$ \citep{Zapata15}.

The relative positional shift of the XZ Tau binary system originated from the orbital motion has also been detected.
On the assumption of the face-on configuration of the orbital plane, \cite{Carrasco09} showed that the circular orbit of XZ Tau B with respect to XZ Tau A reasonably reproduces the detected motions.
On the same assumption, \cite{Krist08} found that the eccentricity of the orbit is very high ($e \sim$0.9),
with the orbital period of $P >$99 yr.
On the other hand, \cite{Hioki09} estimated the eccentricity $e \sim$ 0.5$\pm$0.2 and the period $P = 1010\pm 260$ yr assuming that the orbital plane is perpendicular to the outflow axis.
More recently, \cite{Dodin16} estimated the orbital period of XZ Tau and the eccentricity of the orbit to be $155 < P < 256$ yr and $0.29 < e < 0.64$, respectively.
These results imply that it is not straightforward to derive the 3-dimensional orbital motion from the continuum observations only.
Combined observations of the continuum emission and molecular lines with the velocity information are important to disentangle the ambiguity.

In this paper, we report detection of the misalignment between the two CSDs, orbital motion, and the misalignment between the CSDs and the orbital plane, from the multi-epoch ALMA observations of XZ Tau in the millimeter continuum emission and the $^{12}$CO (2--1) line.
In Section \ref{sec:observations}, the multi-epoch ALMA archival data toward XZ Tau used in this paper are described.
In Section \ref{sec:results}, the results of the 1.3-mm dust-continuum and $^{12}$CO (2--1) emission, and the relative orbital motion of XZ Tau found from the multi-epoch ALMA data are shown.
In Section \ref{sec:discussion}, we show a detailed model of the misaligned disks using RADMC3d \citep{radmc3d}, as well as the orbital solution misaligned from the CSDs.
We discuss these new results in the context of mechanism of binary star formation.
Section \ref{sec:summary} summarizes our main results and discussion.

{\tiny
\begin{table*}
\footnotesize
\centering
\caption{ALMA Archival Data of XZ Tau Adopted for the Present Study}
\label{table:XZ_Tau_property}
\begin{tabular}{lccccc}
\hline
Project name & 2013.1.00105.S & \multicolumn{2}{c}{2016.1.01488.S} & \multicolumn{2}{c}{2017.1.00388.S}\\
PI & Rachel Akeson & \multicolumn{2}{c}{Guillem Anglada} & \multicolumn{2}{c}{Hauyu Baobab Liu}\\
Cycle & Cycle 2 & \multicolumn{2}{c}{Cycle 4} & \multicolumn{2}{c}{Cycle 5}\\
Observing date & 18 Sep. 2015 & 6 Oct. 2016 & 22 Sep. 2017 & 4 Nov. 2017 & 20 Nov. 2017\\
Central frequency & 237.5 GHz & 239.1 GHz & 240.0 GHz & 153.0 GHz & 93.0 GHz\\
Bandwidth & 5.8 GHz & 5.5 GHz & 5.5 GHz & 7.6 GHz & 7.6 GHz\\
Integration time & 121.0 s & 1088.6 s & 2939.3s & 344.7 s & 235.87 s\\
$uv$ range & 31.8 -- 1615.4 $k\lambda$ & 14.3 -- 2461.5 $k\lambda$ & 31.8 -- 9307.7 $k\lambda$ & 107.6 -- 6616.4 $k\lambda$ & 26.3 -- 2442.2 $k\lambda$\\
CASA version used for the calibration & 4.3.1 & 4.7.0 & 4.7.2 & 5.1.1 & 5.1.1\\
\hline
\end{tabular}
\end{table*}
}

{\tiny
\begin{table}
\footnotesize
\centering
\caption{ALMA Archival Data of XZ Tau for the Imaging Analyses}
\label{table:XZ_Tau_List2}
\begin{tabular}{lcc}
\hline
 & 1.3-mm Continuum & $^{12}$CO(2--1)\\
\hline
Project name & \multicolumn{2}{c}{2016.1.01488.S}\\
PI & \multicolumn{2}{c}{Guillem Anglada}\\
Cycle & \multicolumn{2}{c}{Cycle 4}\\
Observing Date & 22 Sep. 2017 & 6 Oct. 2016\\
Weighting & Uniform & Briggs (robust=0.5)\\
Beam size & 0$\farcs$04$\times$0$\farcs$02 & 0$\farcs$19$\times$0$\farcs$14\\
& (P.A.= 16.5$\degr$) & (P.A.= 12.9$\degr$)\\
rms & 0.054 mJy beam$^{-1}$ & 2.96 mJy beam$^{-1}$\\
$\Delta v$ &\nodata & 0.16 km s$^{-1}$\\
\hline
\end{tabular}
\end{table}
}

\section{ALMA data} \label{sec:observations}

{\tiny
\begin{table*}
\footnotesize
\centering
\caption{ALMA Archival Data of XZ Tau Used to Study the Orbital Motion}
\label{table:XZ_Tau_List3}
\begin{tabular}{lccccc}
\hline
& \multicolumn{4}{c}{$uv$ range$>$700 $k\lambda$}\\
\hline
Project name & 2013.1.00105.S & \multicolumn{2}{c}{2016.1.01488.S} & \multicolumn{2}{c}{2017.1.00388.S}\\
PI & Rachel Akeson & \multicolumn{2}{c}{Guillem Anglada} & \multicolumn{2}{c}{Hauyu Baobab Liu}\\
Cycle & Cycle 2 & \multicolumn{2}{c}{Cycle 4} & \multicolumn{2}{c}{Cycle 5}\\
Observing Date & 18 Sep. 2015 & 6 Oct. 2016 & 22 Sep. 2017 & 4 Nov. 2017 & 20 Nov. 2017\\
Weighting & Briggs (robust=1.0) & Briggs (robust=1.0) & Briggs (robust=0.5) & Briggs (robust=1.5) & Briggs (robust=0.5)\\
Beam size & 0$\farcs$13$\times$0$\farcs$10 & 0$\farcs$13$\times$0$\farcs$10 & 0$\farcs$13$\times$0$\farcs$10 & 0$\farcs$13$\times$0$\farcs$10 & 0$\farcs$13$\times$0$\farcs$10\\
 & (P.A.= -3.5$\degr$) & (P.A.= 17.4$\degr$) & (P.A.= -5.4$\degr$) & (P.A.= -14.4$\degr$) & (P.A.= 18.5$\degr$)\\
rms & 0.35 mJy beam$^{-1}$ & 0.19 mJy beam$^{-1}$ & 0.10 mJy beam$^{-1}$ & 0.03 mJy beam$^{-1}$ & 0.15 mJy beam$^{-1}$\\
\hline
\end{tabular}
\end{table*}
}

XZ Tau has been observed with ALMA in several different projects.
In the present paper, we adopt the 230 GHz data of 2013.1.00105.S (PI: Rachel Akeson) and 2016.1.01488.S (PI: Guillem Anglada), and 146 and 86 GHz data of 2017.1.00388.S (PI: Hauyu Baobab Liu).
In the 230 GHz data the $^{12}$CO ($J$=2--1) emission is included along with the 1.3-mm continuum emission.
No apparent line emission is detected in the data of 2017.1.00388.S, and only the 2.1-mm and 3.5-mm continuum data are adopted.
The field center of those archival data matches with the location of XZ Tau.
Table \ref{table:XZ_Tau_property} summarizes the Cycles, observing dates, central frequencies and the bandwidths of the continuum
data, integration time, $uv$ range, and the CASA versions used for the visibility calibrations.
The minimum projected baseline lengths are
14.3 $k\lambda$ at 1.3 mm,
107.6 $k\lambda$ at 2.1 mm,
and 26.3 $k\lambda$ at 3.5 mm.
These minimum projected baseline lengths correspond to the maximum recoverable scales at the 10$\%$ level of $\sim$11$\farcs$5,
$\sim$1$\farcs$5, and $\sim$6$\farcs$3, respectively.


The calibrated visibility data were Fourier-transformed and CLEANed using CASA (version 5.4.1) task tclean to create the continuum and molecular-line images. 
The imaging processes are classified into two categories.
One is the imaging of the 1.3-mm continuum emission and the $^{12}$CO (2--1) emission with the 2016.1.01488.S data.
The purpose of this imaging is to investigate the detailed spatial and velocity structures.
The data taken on 22 Sep. 2017, and the Uniform weighting were adopted to create the 1.3-mm continuum image at the highest angular resolution of $\sim$0$\farcs$03.
While the data taken on the same date and the 2013.1.00105.S data also include the $^{12}$CO (2--1) emission, only the $^{12}$CO (2--1) data of 2016.1.01488.S taken on 6 Oct. 2016 were adopted to create the line images.
This is because we cannot combine the datasets taken on the different dates because of the orbital motion, and because both the sensitivity and angular resolution of the 2016.1.01488.S data are higher than those of the 2013.1.00105.S data.
Briggs weighting with the robust parameter of 0.5 was adopted in the line imaging, which results in the angular resolution of 0$\farcs$19 $\times$ 0$\farcs$14 (P.A. = 12.9$\degr$) and the rms noise level of 3.0 mJy beam$^{-1}$ at a velocity resolution of 0.16 km s$^{-1}$. 
The parameters of the images made are summarized in Table \ref{table:XZ_Tau_List2}.

The other imaging is the continuum-only imaging of all the data. 
These continuum images were made with a restriction of the projected baselines longer than 700 $k\lambda$ and with appropriate Briggs weightings,
to approximately match the resultant synthesized beam sizes and to directly compare the continuum positions at the different epochs.
The imaging parameters and the resultant synthesized beam sizes and the rms noise levels are summarized in Table \ref{table:XZ_Tau_List3}.

\section{results} \label{sec:results}
\subsection{1.3-mm Dust-Continuum Emission} \label{subsec:continuum}

\begin{figure*}
\centering
\includegraphics[width=170mm, angle=0]{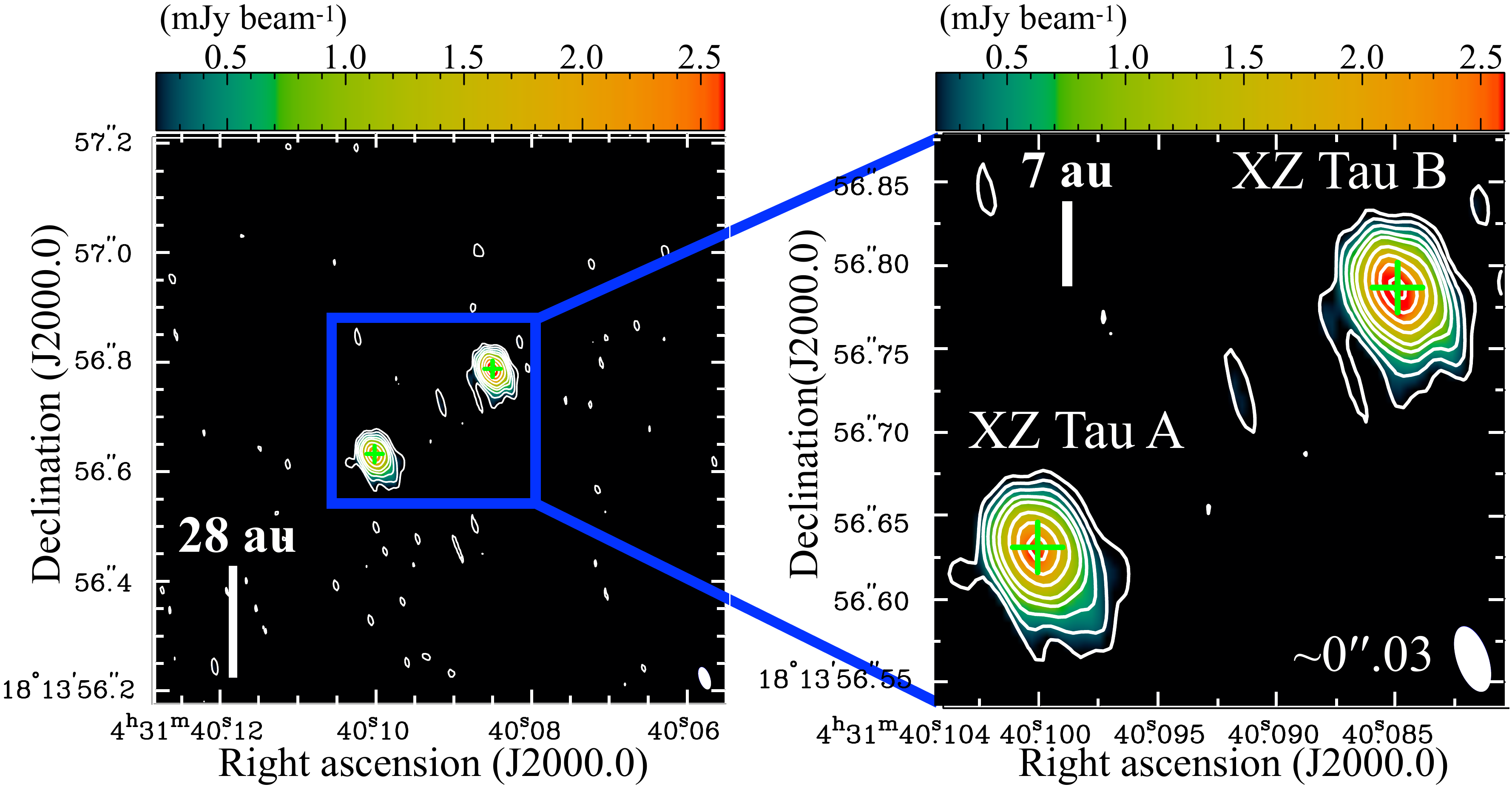}
\caption{
Overall (left panel) and zoom-up (right) views of the 1.3-mm dust-continuum emission of XZ Tau observed with ALMA on 22 Sep. 2017. 
Contour levels are 3$\sigma$, 6$\sigma$, 10$\sigma$, 15$\sigma$, 20$\sigma$, 30$\sigma$, 40$\sigma$, 50$\sigma$, and 60$\sigma$ (1$\sigma$ = 0.05 mJy beam$^{-1}$). 
Green crosses show the centroid positions of the continuum emission obtained from the 2-dimensional Gaussian fitting.
A filled ellipse at the bottom-right corner in each panel shows the synthesized beam (0$\farcs$04 $\times$ 0$\farcs$02; P.A. = 16.5$\degr$).
}
\label{figs_cont}
\end{figure*}

Figure \ref{figs_cont} shows the overall (left panel) and close-up views (right) of the 1.3-mm dust-continuum emission of XZ Tau observed with ALMA on 22 Sep. 2017.
Two 1.3-mm continuum sources located to the southeast and northwest are seen, which most likely trace the CSDs around XZ Tau A and B, respectively \citep{Forgan14}.
In contrast to the ALMA results of the Class I binaries of L1551 NE \citep{takakuwa17} and L1551 IRS 5 \citep{takakuwa20} located in the same L1551 region, no dust-continuum emission surrounding the two CSDs, CBD, is seen toward the Class II binary XZ Tau.
From the two-dimensional Gaussian fitting, the total flux density and the centroid position of the CSD around XZ Tau A are derived to be 5.8 mJy and (04$^{\rm h}$31$^{\rm m}$40.10$^{\rm s}$, 18$^{\rm d}$13$^{\rm m}$56.63$^{\rm s}$), respectively, and those around XZ Tau B 7.3 mJy and (04$^{\rm h}$31$^{\rm m}$40.08$^{\rm s}$, 18$^{\rm d}$13$^{\rm m}$56.79$^{\rm s}$).
Hereafter, we regard these centroid positions as the positions of the individual sources on the observed date.
The continuum sources are barely resolved even at a high angular resolution of $\sim$0$\farcs$03, and the upper limit of the CSD sizes is $\lesssim$15 au.
While the continuum emission are apparently elongated along the northeast to southwest direction, the direction is similar to that of the synthesized beam.
We have also made the continuum images with the other ALMA archival data of XZ Tau, but the images show two point sources.
It is thus not possible to derive the inclination and position angles of these CSDs from the continuum images, and the molecular-line data should be adopted to derive those parameters.

The masses of the individual CSDs ($\equiv M_{\rm d}$) are estimated from their individual continuum flux densities ($\equiv S_\nu$) using the relationship

\begin{equation}
M_{\rm d}=\frac{S_{\nu}d^2}{\kappa_{\nu}B_{\nu}(T_{\rm d})},
\end{equation}

\noindent where $\nu$ is the frequency, $d$ is the distance, $B_\nu(T_{\rm d})$ is the Planck function for dust at a temperature $T_{\rm d}$, and $\kappa_\nu$ is the dust opacity per unit gas + dust mass on the assumption of the gas-to-dust mass ratio of 100.
The canonical relationship,
$\kappa_\nu=\kappa_{\nu_0}(\nu/\nu_0)^\beta$,
and the value of $\kappa_{\rm 250\mu m}$= 0.1 cm$^2$ g$^{-1}$ \citep{Hildebrand83}, and $\beta$ = 1.0 are adopted.
The dust mass opacity at 1.3-mm is then calculated to be $\kappa_{\rm 1.3mm}$ $\sim$ 0.019 cm$^2$ g$^{-1}$.
Regarding the dust temperature, we adopt $T_{\rm d}$ = 10--30 K, which is a typical range of the dust temperatures in Class II disks \citep{akeson19}.
The masses for the CSDs of XZ Tau A and B are calculated to be 
 $\sim$ (0.77--3.47)$\times$10$^{-3}$ $M_{\odot}$
and 
(0.96--4.36)$\times$10$^{-3}$ $M_{\odot}$,
respectively, for $T_{\rm d}$ = 10--30 K.
\cite{akeson19} estimated the dust temperatures and the masses of the disk around XZ Tau A to be 
 $\sim$ 18.8--21.7 K 
and 
 $\sim$ (0.74--1.07)$\times$10$^{-3}$ $M_{\odot}$,
and those XZ Tau B 
 $\sim$ 15.3--19.5 K 
and 
 $\sim$ (1.0--1.5)$\times$10$^{-3}$ $M_{\odot}$. 
Our mass estimates are thus in agreement with those by \cite{akeson19}.

\subsection{$^{12}$CO ($J$=2--1) Emission} \label{subsec:12CO}

\begin{figure*}
\centering
\includegraphics[width=170mm, angle=0]{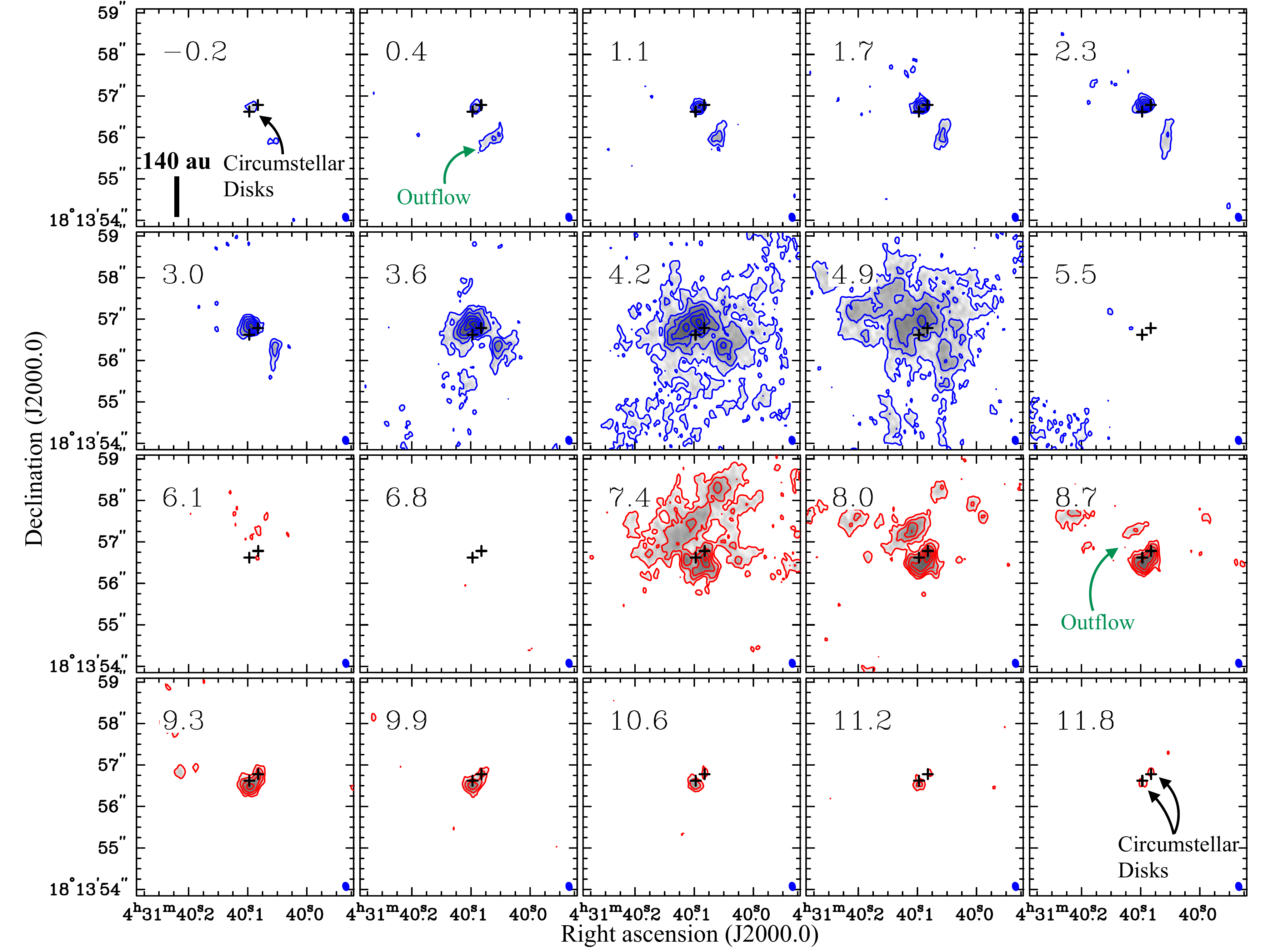}
\caption{
Velocity channel maps of the $^{12}$CO (2--1) emission toward XZ Tau as observed with ALMA on 6 Oct. 2016.
Contour levels are 3$\sigma$, 6$\sigma$, 10$\sigma$, 15$\sigma$, 20$\sigma$, 25$\sigma$, 30$\sigma$, and then in steps of 10$\sigma$ (1$\sigma$ = 3.0 mJy beam$^{-1}$).
Crosses show the positions of the binary.
A filled ellipse at the bottom-right corner shows the synthesized beam (0$\farcs$19 $\times$ 0$\farcs$14; P.A. = 12.9$\degr$).
Numbers in the upper-left corners denote the LSR velocity.
}
\label{figs_12CO_obs}
\end{figure*}

\begin{figure*}
\centering
\includegraphics[width=170mm, angle=0]{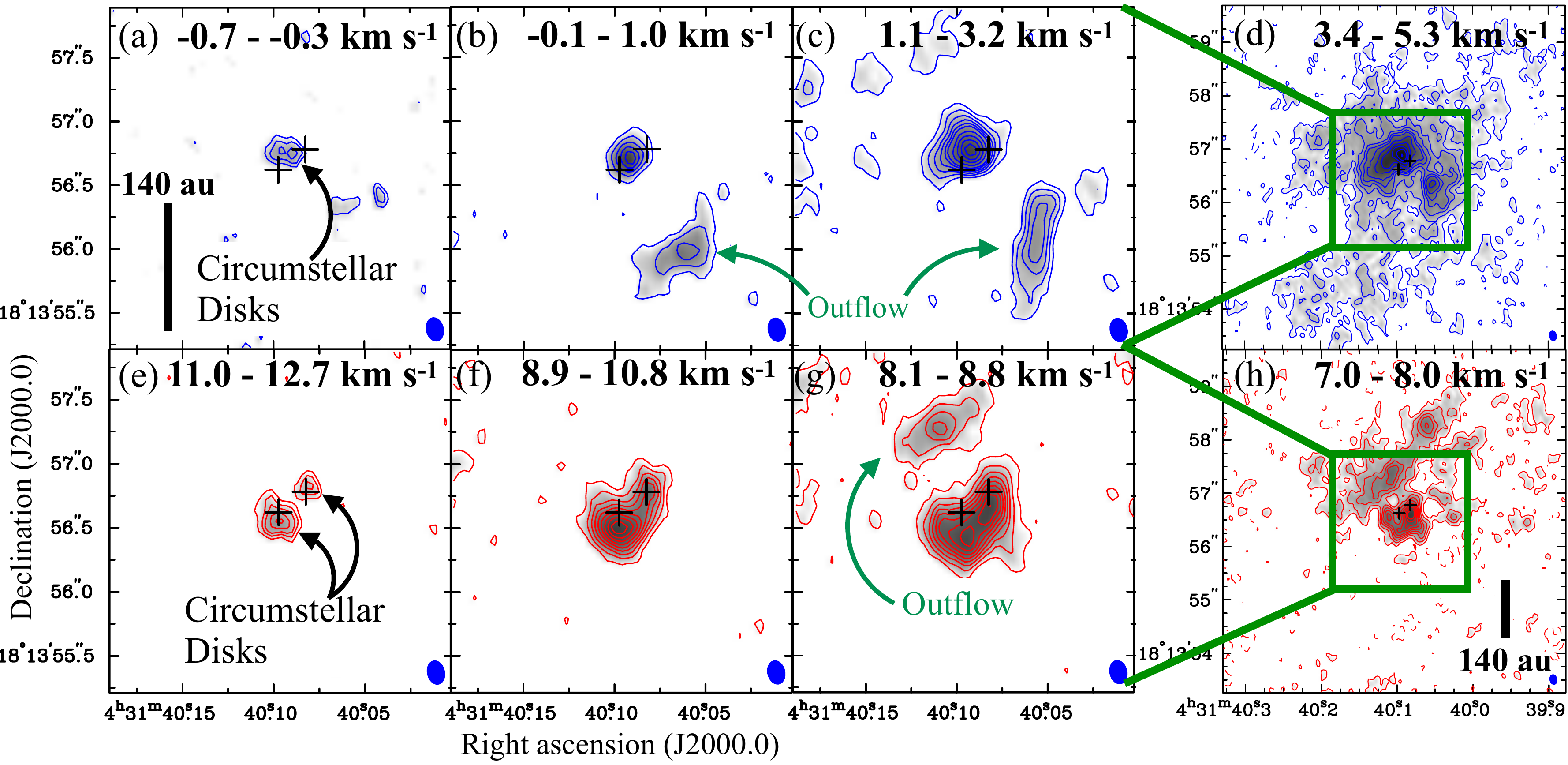}
\caption{
Moment 0 maps of the $^{12}$CO (2--1) emission at representative blueshifted (blue contours) and redshifted (red) velocity ranges in XZ Tau.
The integrated velocity ranges are shown at the top of the relevant panels.
In panel (a) contour levels are 3.8$\sigma$, 5.3$\sigma$, and 6.8$\sigma$ (1$\sigma$ = 1.0 mJy beam$^{-1}$ km s$^{-1}$).
In panel (b) contour levels are in steps of 5$\sigma$ (1$\sigma$ = 1.58 mJy beam$^{-1}$ km s$^{-1}$).
In panel (c) contour levels starts from 5$\sigma$ in steps of 5$\sigma$ until 30$\sigma$, and then 40$\sigma$, 50$\sigma$, 60$\sigma$, and 70$\sigma$ (1$\sigma$ = 1.58 mJy beam$^{-1}$ km s$^{-1}$).
In panel (d) contour levels starts from 5$\sigma$ in steps of 5$\sigma$ until 30$\sigma$, and then 40$\sigma$, 50$\sigma$, 60$\sigma$, 70$\sigma$, and 80$\sigma$ (1$\sigma$ = 1.64 mJy beam$^{-1}$ km s$^{-1}$).
In panels (e) and (f) contour levels are in steps of 5$\sigma$ (1$\sigma$ = 1.06 mJy beam$^{-1}$ km s$^{-1}$).
In panel (g) contour levels are in steps of 5$\sigma$ (1$\sigma$ = 1.58 mJy beam$^{-1}$ km s$^{-1}$).
In panel (h) contour levels are in steps of 5$\sigma$ (1$\sigma$ = 1.26 mJy beam$^{-1}$ km s$^{-1}$).
Crosses show the positions of the binary, and a filled ellipse at the bottom-right corner shows the synthesized beam (0$\farcs$19 $\times$ 0$\farcs$14; P.A. = 12.9$\degr$).
}
\label{figs_12COmom_obs}
\end{figure*}

\begin{figure*}
\centering
\includegraphics[width=170mm, angle=0]{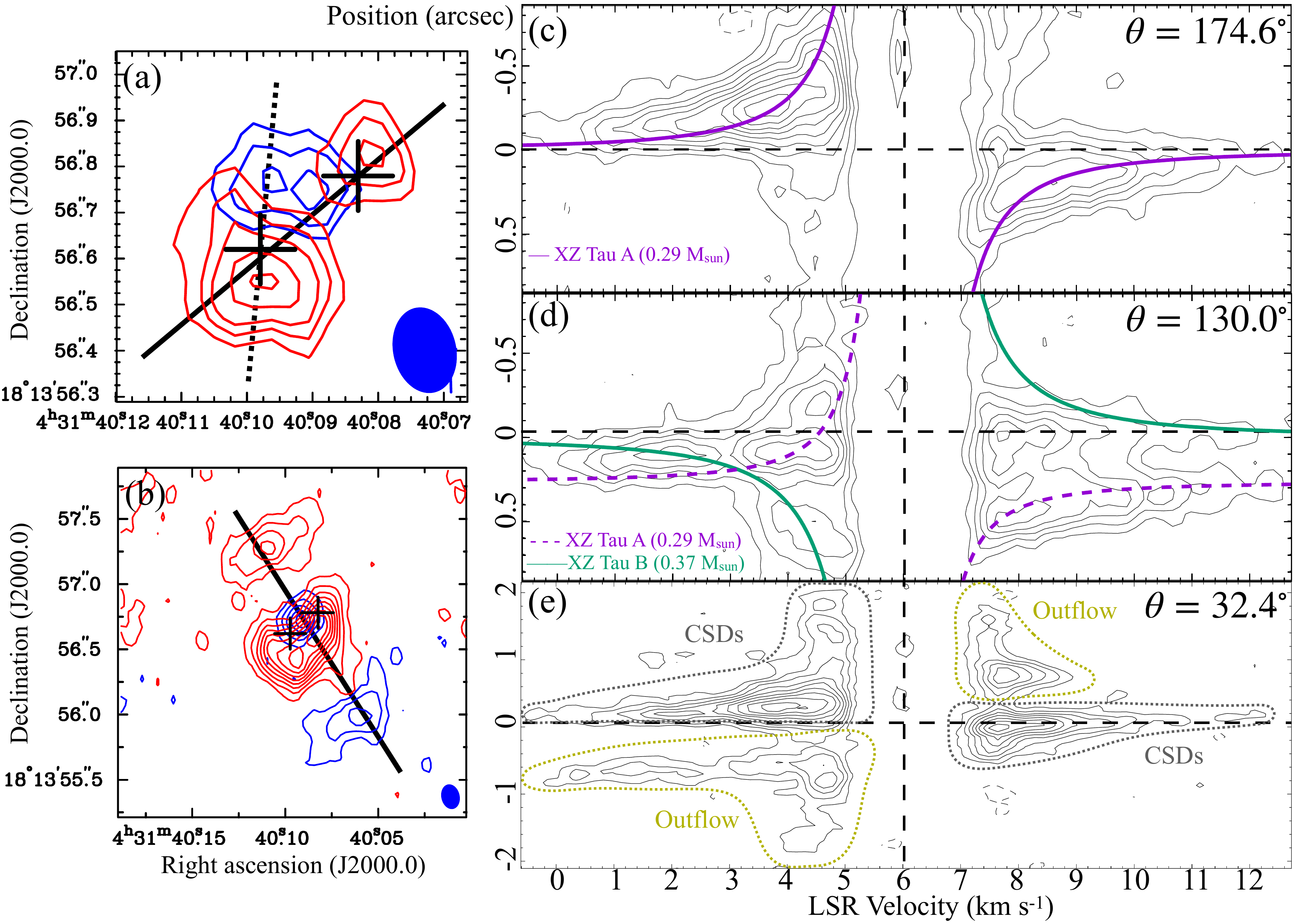}
\caption{
Maps of the blueshifted (blue contour) and redshifted (red) $^{12}$CO (2–-1) emission (panels a, b), and position-velocity (P-V) diagrams of the $^{12}$CO (2--1) emission (c, d, e).
In panel a), the integrated velocity ranges of the blue- and redshifted emission are -0.7--0.3 km s$^{-1}$ and 11.0--12.7 km s$^{-1}$, respectively, and those in panel b) -0.1--1.0 km s$^{-1}$ and 8.1--8.8 km s$^{-1}$.
In panels a) and b), crosses show the positions of the binary, and filled ellipses at the bottom-right corners the synthesized beam (0$\arcsec$.19 $\times$ 0$\arcsec$.14; P.A. = 12.9$\degr$).
Black dashed and solid lines in panel a) denote the major axes of the circumstellar disks of XZ Tau A and B, and the cut lines of the P-V diagrams in panels c) and d), respectively.
A black solid line in panel b) denotes the cut line of the P-V diagram in panel e).
In panel c), d), and e), horizontal dashed lines denote the positions of XZ Tau A, B, and the middle position between the binary, respectively, while the vertical dashed lines the systemic velocity of 6.0 km s$^{-1}$.
Purple solid and dashed curves show Keplerian rotations of the circumstellar disks of XZ Tau A, and green solid curves that of XZ Tau B, with the adopted inclination angle of 55$\degr$.
In panel a), contour levels of the blueshifted emission are 3.8, 5.3, and 6.8$\sigma$ (1$\sigma$ = 1.0 mJy beam$^{-1}$ km s$^{-1}$), and those of the redshifted emission in steps of 5$\sigma$ (1$\sigma$ = 1.06 mJy beam$^{-1}$ km s$^{-1}$).
Contour levels in panel b) are in steps of 5$\sigma$ (1$\sigma$ = 1.58 mJy beam$^{-1}$ km s$^{-1}$).
In panels (c), (d), and (e) contour levels are in steps of 3$\sigma$ (1$\sigma$ = 3.0 mJy beam$^{-1}$).
}
\label{figs_pv_obs}
\end{figure*}

Figure \ref{figs_12CO_obs} shows velocity channel maps of the $^{12}$CO (2--1) emission as observed on 6 Oct. 2016. While the $^{12}$CO (2--1) emission is also included and detected in the Cycle 2 data taken on 18 Sep. 2015, in the present paper we only adopt the 2016 data because of its higher sensitivity and angular resolution.
The systemic velocity of XZ Tau in LSR has been measured to be $v_{\rm sys} \simeq$ 6.0 km s$^{-1}$ \citep{ALMA15,Zapata15}.
While this systemic velocity appears to be consistent with Figure \ref{figs_12CO_obs}, around the systemic velocity the $^{12}$CO (2--1) emission is singnificantly suppressed due to the effect of the missing flux.
Figure \ref{figs_12CO_obs} shows that in the blueshifted velocity range, there are primarily two components, one located between the binary and the other to the southwest of the binary. The peak position of the former component gradually shifts toward the northeast from $V_{\rm LSR}$ = 1.7 km s$^{-1}$.
On the other hand, in the redshifted velocity range,
two, apparently distinct emission components associated with the individual binary stars are seen, plus another redshifted emission to the northeast of the binary.

To investigate these features in more detail,
in Figure \ref{figs_12COmom_obs} we show the $^{12}$CO (2--1) maps integrated over the different velocity ranges; very high-, high-, middle-, and low-velocity blueshifted (upper panels) and redshifted (lower panels) ranges.
Note that the maps in the very high velocities include velocity channels outside those shown in Figure \ref{figs_12CO_obs}, because the integrations over these additional channels unveil the characteristic features more clearly.
In the highest blueshifted velocity, the emission component located between XZ Tau A and B as seen in Figure \ref{figs_12CO_obs}
shows two peaks, 
one located to the north of XZ Tau A and the other to the southeast of XZ Tau B (Figure \ref{figs_12COmom_obs}a).
In the highest redshifted velocity, on the other hand, two distinct emission components, one located to the south of XZ Tau A and the other to the northwest of XZ Tau B, are present (Figure \ref{figs_12COmom_obs}e).
Toward the lower blueshifted velocities, the peak position of the blueshifted emission gradually shifts to the northeast, and another blueshifted emission to the southwest appears, as seen in Figure \ref{figs_12CO_obs} (Figure \ref{figs_12COmom_obs}b and c).
In the lower redshifted velocities, the two distinct redshifted emission components become larger, and they connect at the southwest of the binary (Figure \ref{figs_12COmom_obs}f and g).
Another redshifted component to the north of the binary, as found in Figure \ref{figs_12CO_obs}, is also seen (Figure \ref{figs_12COmom_obs}g).
In the lowest blueshifted and redshifted ranges (Figure \ref{figs_12COmom_obs}d and h), the $^{12}$CO emission extends in the entire region,
but the directions of the peak positions of these detected components are unchanged.

In Figure \ref{figs_pv_obs}a we show a zoomed view of the highest-velocity blueshifted and redshifted $^{12}$CO (2--1) emission.
The blueshifted emission located between the binary is separated into two emission peaks, one to the north of XZ Tau A and the other to the southeast of XZ Tau B.
The redshifted counterparts are seen to the south of XZ Tau A and the northwest of XZ Tau B. 
Figure \ref{figs_pv_obs}c shows the Position-Velocity (P-V) diagram along the north-south direction passing through XZ Tau A (dashed line in Figure \ref{figs_pv_obs}a). 
The position angle of the P-V cut, $\theta$ = 174.6$\degr$, is chosen so that the line passes through the northern blueshifted and southern redshifted peaks around XZ Tau A.
The velocities of northern blueshifted and southern redshifted
components appear to be higher as the position becomes closer to the stellar position.
These results suggest presence of Keplerian rotation, and the $^{12}$CO (2--1) emission located to the north and south of XZ Tau A likely trace the CSD around XZ Tau A. 
The P-V diagram along the northwest to southeast direction passing through XZ Tau B (solid line in Figure \ref{figs_pv_obs}a) is shown in Figure \ref{figs_pv_obs}d.
The position angle of this P-V cut ($\theta$ = 130.0$\degr$) is chosen to pass through the southeastern blueshifted and northwestern redshifted peaks around XZ Tau B.
In this P-V diagram the zero position is set to be the continuum centroid position of XZ Tau B.
On the redshifted side, there are two emission protrusions sticking to the higher redshifted velocities, one located to the northwest of XZ Tau B and the other to the southeast. 
The southeastern redshifted component corresponds to the redshifted part of the CSD around XZ Tau A, as seen in Figure \ref{figs_pv_obs}c.
On the other hand, the redshifted emission to the northwest of XZ Tau B is distinct from this component, and likely traces the CSD around XZ Tau B. 
On the blueshifted side, there appears a blueshifted component to the southeast of XZ Tau B with a Keplerian rotation signature. 
This component appears to be the blueshifted counterpart to the redshifted emission to the northwest of XZ Tau B, and these components likely comprise the Keplerian rotation in the CSD around XZ Tau B.

To verify whether the position angles of the P-V cuts shown in Figure \ref{figs_pv_obs}a correspond to the rotational directions of the CSDs or not, we made a number of P-V diagrams passing through the binary stars by varying the position angles of the P-V cuts. 
The ranges of the position angles are $\pm$20$\degr$ from the central values, and the step is 5 degree. We investigated all of these P-V diagrams and verified that the original position angles show the highest velocity components and Keplerian rotation signatures most clearly.
We here define $\theta$ = -5.4$\degr$ and 130.0$\degr$ as the major axes of the CSDs around XZ Tau A and B, respectively. 
The step of the P-V search of 5$\degr$ can be regarded as the error of the position angles.
We have also attempted Markov Chain Monte Carlo (MCMC) fitting of geometrically-thin Keplerian disk models to the observed $^{12}$CO channel maps to derive the best-fit position angles and the errors.
The fitting parameters are central stellar masses, position and inclination angles of the disks, R.A. and Dec. of the stellar positions, and systemic velocities for both XZ Tau A and B.
We found, however, that MCMC never converges but always diverges from certain steps.
Fixing the stellar positions to the continuum positions or the stellar masses to the published values \citep{Hartigan03} does not help.
The failure of the MCMC fitting could be due to the effect of the missing fluxes around ($\sim\pm$1 km s$^{-1}$) the inferred systemic velocity of 6 km s$^{-1}$ (see Figures \ref{figs_12CO_obs},  \ref{figs_pv_obs}c, d, e).
Then we just adopt the position angles derived from the manual inspection of the P-V diagrams.

From the optical and near-infrared spectroscopic studies the masses of XZ Tau A and B have been measured to be 0.29 $M_{\odot}$ and 0.37 $M_{\odot}$, respectively \citep{Hartigan03}.
With these central stellar masses Keplerian rotation curves are drawn in Figure \ref{figs_pv_obs}c and d. 
The inclination angles are chosen so that the Keplerian rotation curves approximately delineate the emission ridges of the components in the P-V diagrams. 
The adopted inclination angles of the CSDs in XZ Tau A and B are both 55.0$\degr$.
Note that in section \ref{sec:discussion} we will also show our modeling effort to derive the values of the inclination angles.
We also note that the adopted centroid velocity of the Keplerian rotations ($V_{\rm LSR}$ = 6.0 km s$^{-1}$) is identical between XZ Tau A and B, and that different centroid velocities of the CSDs are not required to approximately reproduce the Keplerian rotations.

Figure \ref{figs_pv_obs}b shows a picture of the blueshifted and redshifted $^{12}$CO (2--1) emission at intermediate velocity ranges (-0.1 km s$^{-1}$ to 1.0 km s$^{-1}$ and 8.1 km s$^{-1}$ to 8.8 km s$^{-1}$).
There are another blueshifted and redshifted emission components to the southwest and northeast of the binary, respectively.
The P-V diagram along the northeast to southwest direction, passing through the northeastern redshifted and southwestern blueshifted $^{12}$CO emission peaks (solid line in Figure \ref{figs_pv_obs}b), is shown in Figure \ref{figs_pv_obs}e.
In this P-V diagram the zero position is set to be the middle position between XZ Tau A and B.
Toward the southwest, the velocity of the blueshifted emission becomes higher as the position is away from the stellar locations.
The redshifted counterpart of this blueshifted component is seen to the northeast.
In XZ Tau a bipolar molecular outflow has been detected \citep{Zapata15}, and the blue- and redshifted components of the outflow are located to the southwest and northeast, respectively.
These results imply that the southwestern blue- and northeastern redshifted $^{12}$CO emission components most likely trace the associated molecular outflow.

\subsection{Orbital motion} \label{subsec:orbital_motion}

\begin{figure*}
\centering
\includegraphics[width=170mm, angle=0]{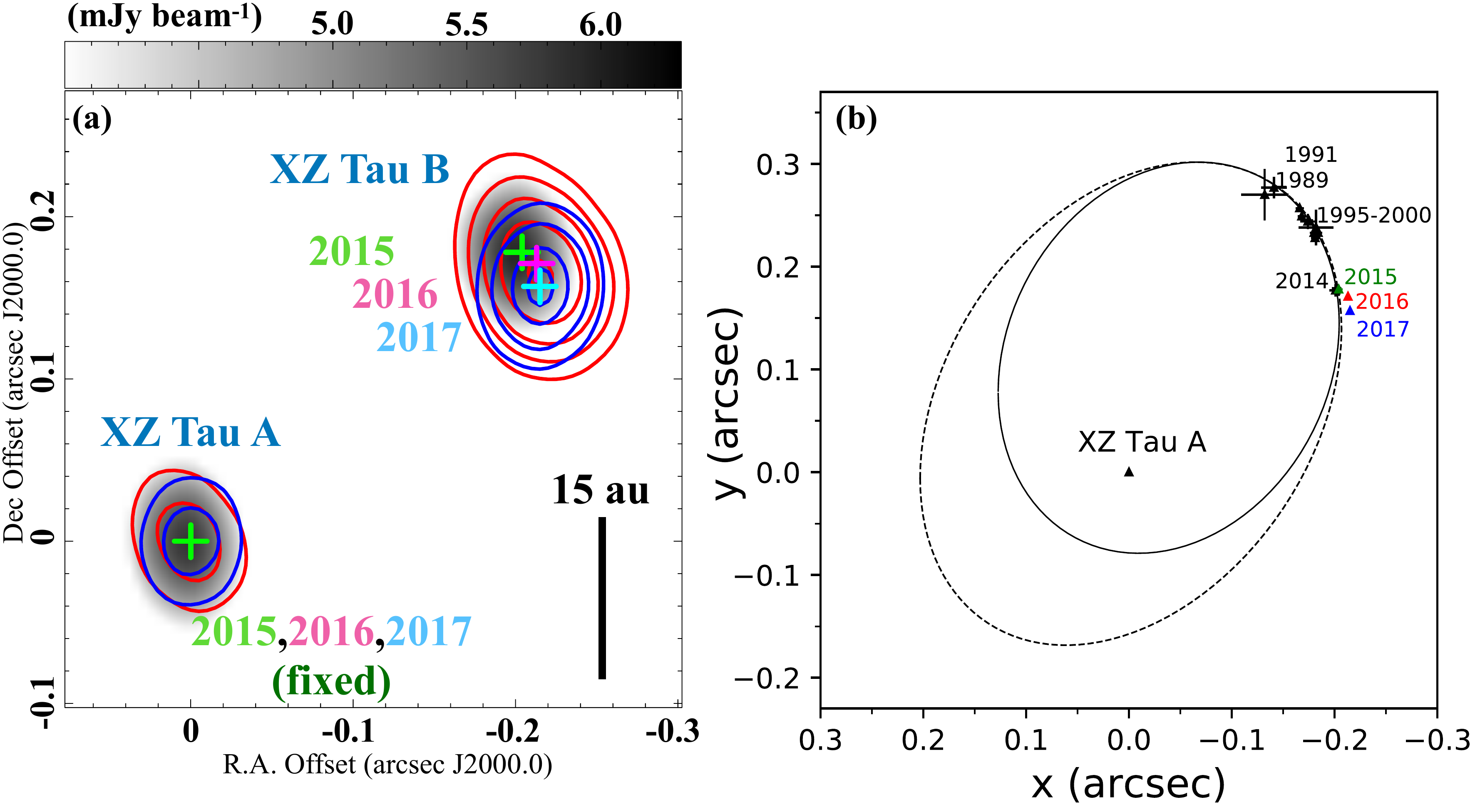}
\caption{
(a) Comparison of the 1.3-mm dust-continuum images of XZ Tau observed in 2015 (gray), 2016 (red contour), and 2017 (blue contour) with ALMA.
Contour levels of the 2016 and 2017 images are 19$\sigma$, 23$\sigma$, 27$\sigma$, 31$\sigma$, 35$\sigma$, 30$\sigma$ (1$\sigma$ = 0.02 mJy beam$^{-1}$), and 50$\sigma$, 60$\sigma$, 70$\sigma$, 75$\sigma$ (1$\sigma$ = 0.01 mJy beam$^{-1}$), respectively.
Green, red and blue crosses show the centroid positions of the continuum emission observed in 2015, 2016, and 2017, respectively.
The three images taken at the three different epochs are shifted to have the same origin at the location of XZ Tau A.
(b) Relative orbital motion of XZ Tau B with respect to XZ Tau A, derived from the multi-epoch NIR observations \citep{Dodin16} along with the present ALMA continuum data.
The centroid positions shown in panel a) are overplotted on Figure 1 of \cite{Dodin16}. The error bars of the ALMA positions are smaller than the size of the markers.
}
\label{figs_orbital_motion}
\end{figure*}

We have also made the 1.3-mm continuum images of XZ Tau at the three different observing epochs (18 Sep. 2015, 6 Oct. 2016, and 22 Sep. 2017; see Table \ref{table:XZ_Tau_List3}) at a common angular resolution of 0$\farcs$13 $\times$ 0$\farcs$10 ($\sim$ 19 au $\times$ 15 au) by adjusting the Briggs parameters. 
{\tiny
\begin{table*}
\footnotesize
\centering
\caption{Detected Proper Motion of XZ Tau B with respect to XZ Tau A }
\label{table:Proper_Motion}
\begin{tabular}{cccccccc}
\hline
Observing Date & R.A. Offset\tablenotemark{a} & Dec Offset\tablenotemark{a} & Offset Date & $\Delta\alpha$\tablenotemark{b} & $\Delta\delta$\tablenotemark{b} & $v_{\alpha}$\tablenotemark{c}  & $v_{\delta}$\tablenotemark{c}\\
& [arcsec] & [arcsec] & [days] & [au] & [au] & [au\ yr$^{-1}$] & [au\ yr$^{-1}$]\\
\hline
18 Sep. 2015 & -0.204 {$\pm$} 0.010 & 0.178 {$\pm$} 0.010 & 0 & 0 & 0 & \multirow{2}{*}{-1.2 {$\pm$} 1.9} & \multirow{2}{*}{-1.0 {$\pm$} 2.1}\\
6 Oct. 2016 & -0.213 {$\pm$} 0.004 & 0.171 {$\pm$} 0.004 & 384 & -1.3 {$\pm$} 2.0 & -1.0 {$\pm$} 2.0 & \multirow{2}{*}{-0.3 {$\pm$} 1.0} & \multirow{2}{*}{-2.1 {$\pm$} 1.0}\\
22 Sep. 2017 & -0.215 {$\pm$} 0.003 & 0.157 {$\pm$} 0.003 & 351 & -0.3 {$\pm$} 1.0 & -2.0 {$\pm$} 1.0 & &\\
\hline
\end{tabular}
\vspace{1mm}
\begin{flushleft}
a. Position of XZ-Tau B with respect to that of XZ Tau A in each epoch.\\
b. Relative shift of the XZ Tau B position from that of the last epoch.\\
c. 2-dimensional transverse velocity between the epochs.\\
\end{flushleft}
\end{table*}
}
To extract a possible orbital motion of the binary stars without any uncertainty of the global proper motion of the sky, we derived the relative positions of XZ Tau B with respect to the location of XZ Tau A over the three epochs. Figure \ref{figs_orbital_motion}a demonstrates the detected relative motions.
There is a systematic positional shift of XZ Tau B relative to XZ Tau A, and the position of XZ Tau B is moving toward the southwest.

To investigate whether the detected positional shifts are real or not, we have also made the images of the 2.1-mm and 3.5-mm data taken on 4 Nov. and 20 Nov. 2017, respectively, and compared the continuum positions to the 1.3-mm continuum positions on 22 Sep. 2017.
Assuming that these three 2017 data taken within two months trace the same position, we estimated the absolute positional uncertainty of the ALMA observations to be $\sim$ 3.0 mas.
We have also calculated the astrometric uncertainty following the ALMA knowledge base
\footnote{https://help.almascience.org/kb/articles/what-is-the-astrometric-accuracy-of-alma}.
We found that, for our data, a limit of 3 mas is always the largest among the three estimates described in the knowledge base, and thus this must be the astrometric error of our data, consistent with the above estimate.
If this 3 mas error applies to both the XZ Tau A and B positions independently, the error of the position of XZ Tau B with respect to that of XZ Tau A should be $\sqrt{3^2+3^2} \sim$ 4.2 mas.
In reality, the phase variations between the two closely-located positions are likely much smaller, and this should be regarded as a conservative estimate.
The observed positional shift from 2015 to 2017 is significant with respect to this error, and we conclude that the observed positional shifts over the three observing epochs are real.
The detected shift from 2015 to 2017 is $\Delta$R.A. $\sim$ -1.6 au and $\Delta$Dec $\sim$ -3.0 au.
The derived positions and the errors, the time separations, and the corresponding 2-dimensional velocities are summarized in Table \ref{table:Proper_Motion}.

In Figure \ref{figs_orbital_motion}b, the positional shifts as observed with ALMA are compared to the previously identified orbital motion of XZ Tau from the NIR observations \citep{Dodin16}.
The ALMA positions are in good agreement with the overall orbital motion of the binary starting from 1989.
We thus suggest that the present ALMA archival data trace the orbital motion of the binary, and add three more years $i.e.$, 2015 to 2017, to the previous NIR observations in 1989--2014.

{\tiny
\begin{table*}
\footnotesize
\centering
\caption{Model Parameters}
\label{table:model_parameters}
\begin{tabular}{lcc}
\hline
Parameter & \multicolumn{2}{c}{Value}\\
    & XZ Tau A & XZ Tau B\\
\hline
\multicolumn{3}{c}{Fixed parameter}\\
\hline
Inner cutoff radius of the disk & \multicolumn{2}{c}{0.01 au}\\
Outer cutoff radius of the gas disk & \multicolumn{2}{c}{260 au}\\
$r_0$ & \multicolumn{2}{c}{10 au}\\
Elevation range of the model domain &  \multicolumn{2}{c}{$\leq$ 40.1 $\degr$}\\
$X$($^{12}$CO) & \multicolumn{2}{c}{9.5 $\times$ 10$^{-5}$}\\
Gas to dust mass ratio & \multicolumn{2}{c}{100}\\
H$_2$ ortho to para ratio & \multicolumn{2}{c}{3.0}\\
Turbulence velocity & \multicolumn{2}{c}{0.4 km s$^{-1}$}\\
Stellar mass & 0.29 $M_\odot$ & 0.37 $M_\odot$\\
Luminosity & 0.31 $L_\odot$ & 0.17 $L_\odot$\\
\hline
\multicolumn{3}{c}{Varied parameter}\\
\hline
Dust surface density at $r_0$ & 90$^{+14}_{-30}$ g cm$^{-2}$ & 90$^{+12}_{-18}$ g cm$^{-2}$\\
Outer cutoff radius of the dust disk & 6.0$^{+0.6}_{-0.4}$ au & 7.0$^{+0.6}_{-0.2}$ au\\
Scaling factor of the disk scale height $\lambda$ & 1.3$\pm$0.1 & 1.5$\pm$0.2\\
Disk P.A. & -5.4$\pm$5.0$\degr$ & 130.0$\pm$5.0$\degr$\\
Disk inclination angle $i$ & 68.0$\pm$2.0 $\degr$ & 64.0$\pm$4.0 $\degr$\\
\hline
\end{tabular}
\end{table*}
}

\section{Discussion} \label{sec:discussion}

\begin{figure*}
\centering
\includegraphics[width=170mm, angle=0]{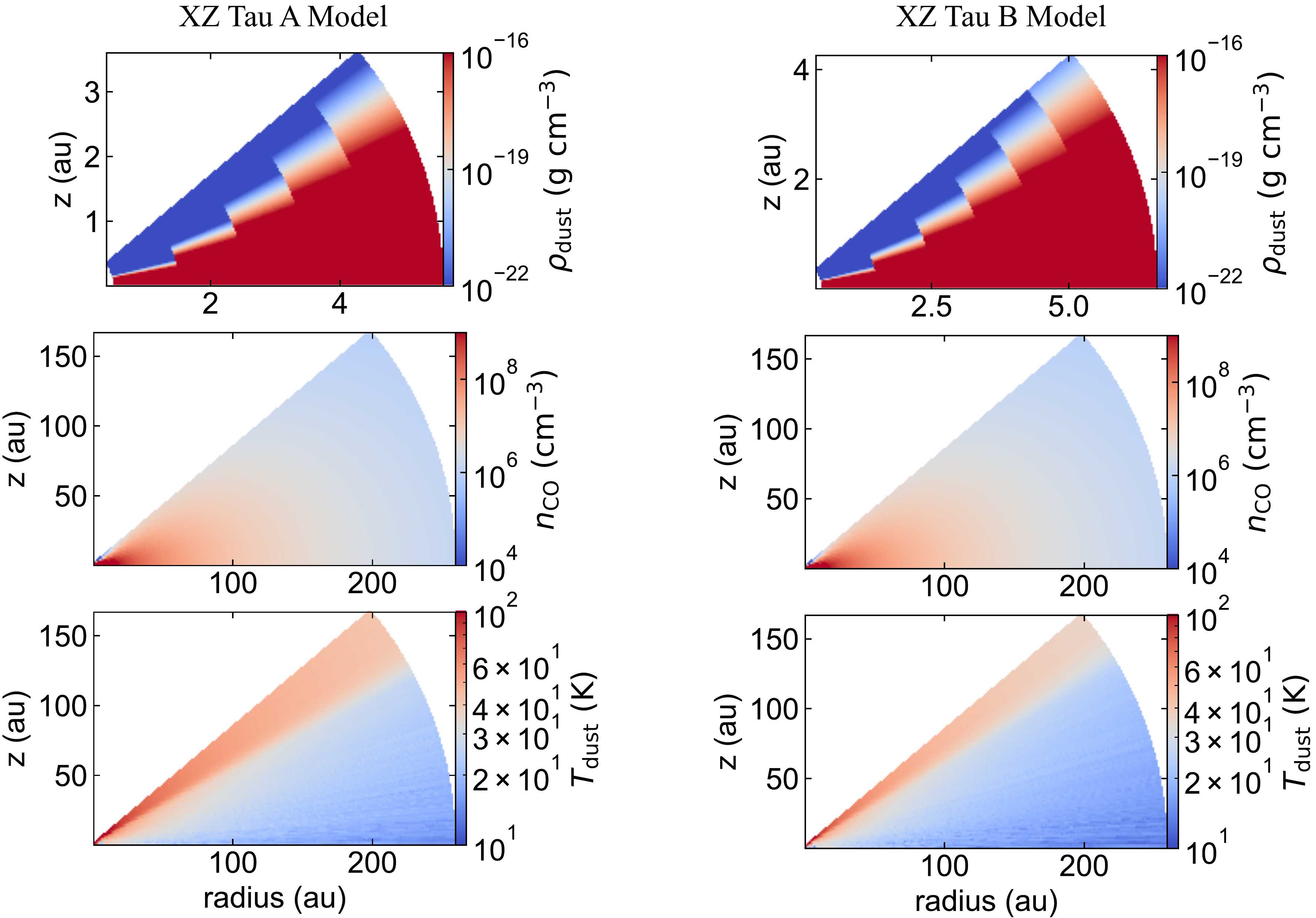}
\caption{
Model distributions of the dust density (top panel), $^{12}$CO number density (middle), and the dust temperature (bottom) of XZ Tau A (left panel) and B (right panel).
}
\label{fig:model_parameter}
\end{figure*}

\begin{figure*}
\centering
\includegraphics[width=160mm, angle=0]{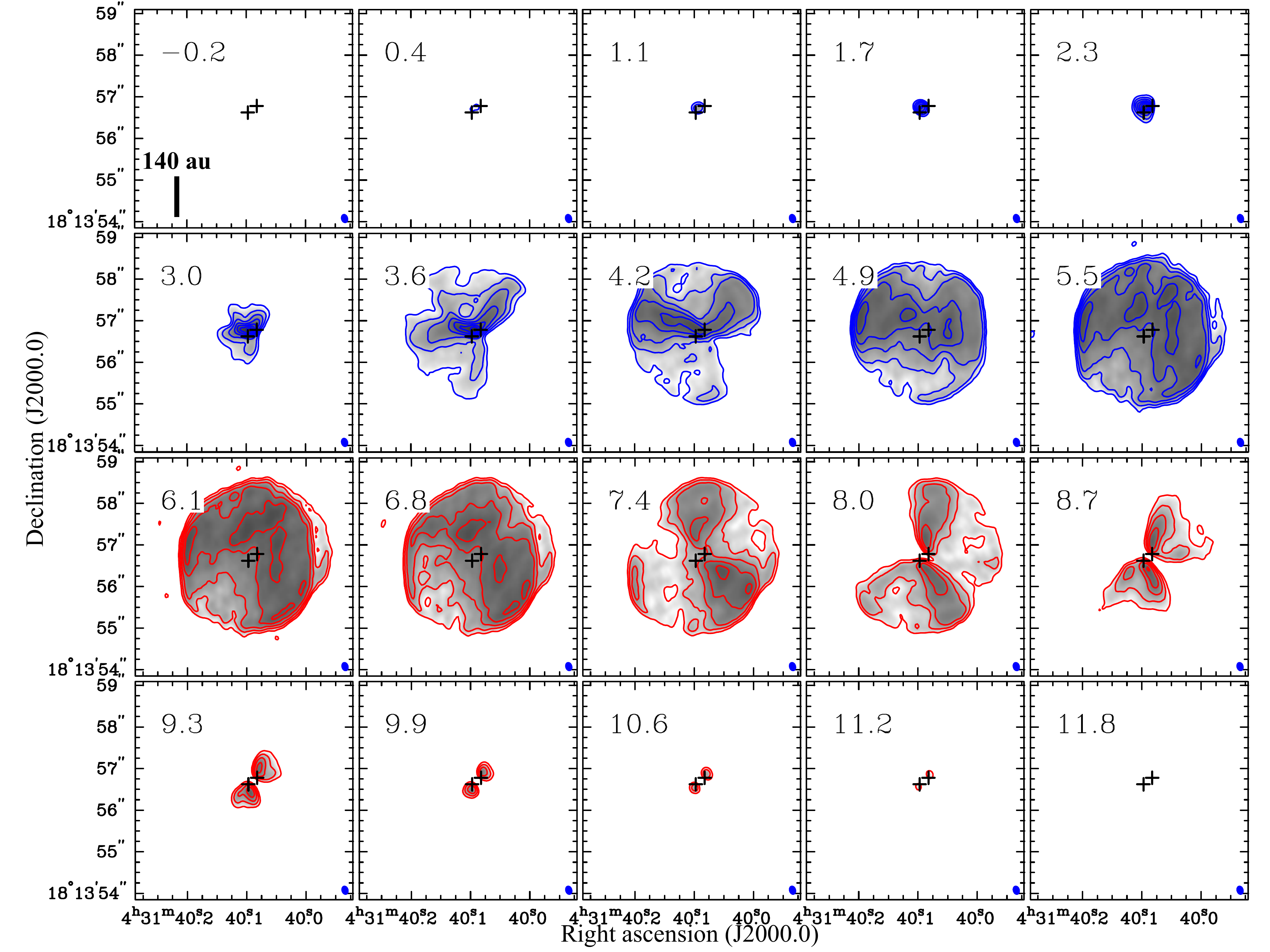}
\caption{
Model velocity channel maps of the $^{12}$CO (2-1) emission. 
Contour levels and symbols are the same as those in Figure \ref{figs_12CO_obs}. 
}
\label{figs_12CO_model}
\end{figure*}

\begin{figure*}
\centering
\includegraphics[width=160mm, angle=0]{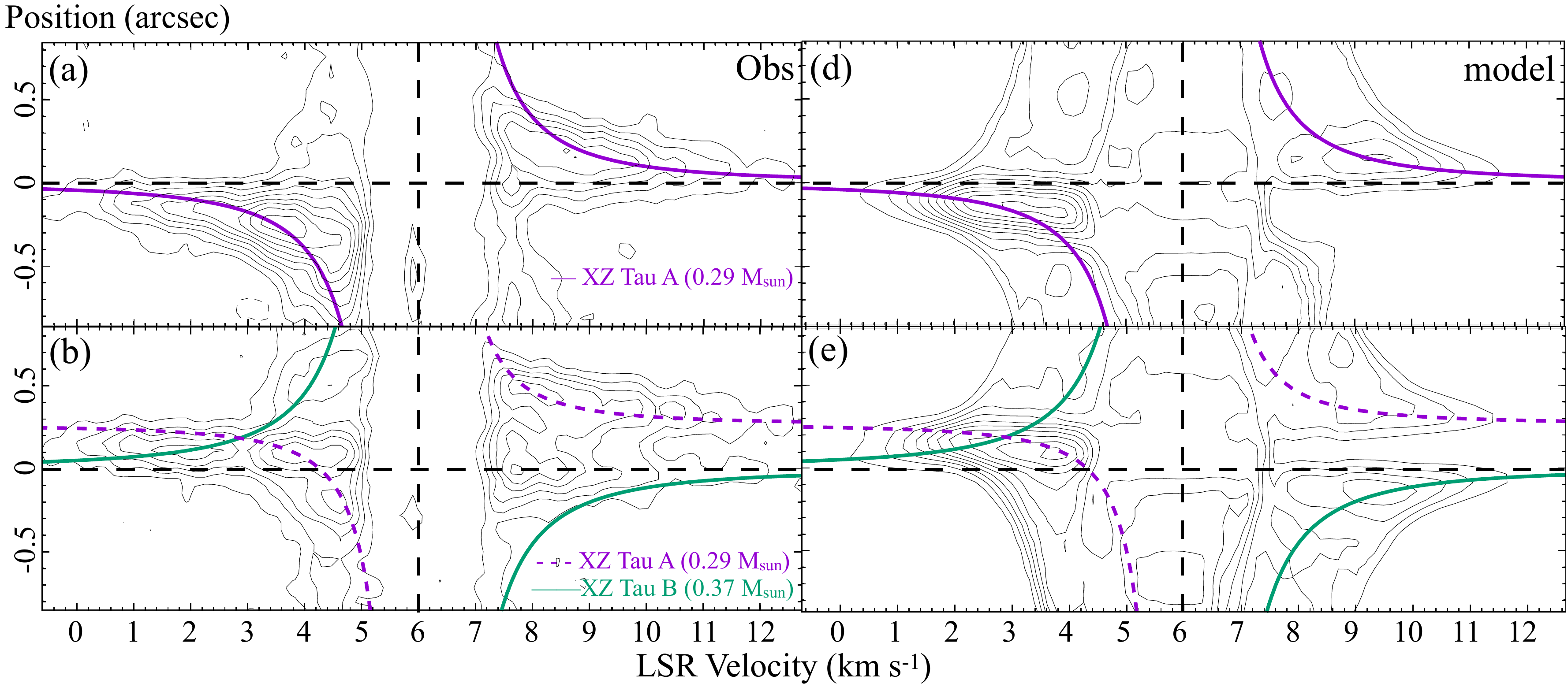}
\caption{
Observed (left panels) and model (right) P-V diagrams of the $^{12}$CO (2--1) emission along the major axes of the circumstellar disks of XZ Tau A (upper panels) and B (lower).
Contour levels are in steps of 3$\sigma$ (1$\sigma$= 3.0 mJy beam$^{-1}$).
Purple solid and dashed curves show Keplerian rotations in the circumstellar disks around XZ Tau A with the inclination angle of 68$\degr$, and green solid curves those around XZ Tau B with the inclination angle of 64$\degr$. 
}
\label{figs_pv_model}
\end{figure*}

\subsection{Modeling of the Misaligned CSDs} \label{kinematics_disk}

ALMA observations of the binary system XZ Tau have found compact, unresolved dusty CSDs in the 1.3-mm continuum emission, whereas in the $^{12}$CO (2--1) emission ALMA observations have unveiled resolved velocity structures in the CSDs.
Toward XZ Tau A, the blue- and redshifted $^{12}$CO emission are located to north and south, respectively.
On the other hand, toward XZ Tau B the blue- and redshifted emission are located to the southeast and northwest.
These results suggest that the two CSDs around the individual binary stars are misaligned with each other.
In addition, there are extended blue- and redshifted $^{12}$CO (2--1) emission located to the southwest and northeast, which could be attributed to the associated molecular outflow as already reported by \cite{Zapata15}.

To interpret the spatial and velocity structures of the CSDs around the binary components in more detail, we constructed 2.5-dimensional, axisymmetric disk models.
The density distributions of the disk materials are set as follows.
The vertical scale height $H(r)$ is assumed to be,
\begin{equation}
    H (r) = \lambda \frac{c_{\rm s}}{\Omega_{\rm K}},
\end{equation}
where 
\begin{equation}
    c_{\rm s} = \sqrt{\frac{k_{\rm B} T_0}{m_{\rm p} \mu}},
\end{equation}
\begin{equation}
    \Omega_{\rm K} = \sqrt{\frac{G M_{\star}}{r^3}}
\end{equation}
are the sound speed and the Keplerian angular velocity, respectively, and $k_{\rm B}$ denotes the Boltzmann constant, $T_0$ the midplane dust temperature at $r$ = $r_0$, $m_{\rm p}$ the proton mass, $\mu$ (=2.33) the mean molecular weight, $G$ the gravitational constant, and $M_{\star}$ denotes the stellar mass.
$r_0$ is fixed to be 10 au.
The coefficient $\lambda$ is the control parameter to adjust the scale height with respect to the vertical hydrostatic equilibrium. 
For simplicity, the midplane temperature $T_{0}$ is assumed to be uniform and fixed to be 80 K. 
Note that this temperature is adopted simply to calculate the scale height and the density distributions of the materials.
A true gas and dust kinetic temperature will be calculated at a later process with a given density distribution.
Then both $T_{0}$ and $\lambda$ are degenerate, since both parameters are just adopted to calculate the scale height.

The dust and gas surface density profile $\Sigma(r)$ is assumed to follow the power-law profile as
\begin{equation}
    \Sigma (r) = \Sigma_{0} \left(\frac{r}{r_0}\right)^p,
\end{equation}
where $\Sigma_{0}$ is the surface density at $r$ = $r_0$, and $p$ is the power-law index.
Then the volume density of the materials $\rho$ can be described as
\begin{equation}
    \rho(r, z) = \frac{\Sigma(r)}{H(r)\sqrt{2\pi}}\ exp\left( {\frac{-z^2}{2H(r)^2}} \right).
\end{equation}

As the disk model requires a number of free parameters, we first reduce the number of the free parameters as follows.
The stellar masses and luminosities of XZ Tau A and B are adopted to be $M_{\rm A}$ = 0.29 $M_{\sun}$, $L_{\rm A}$ = 0.31 $L_{\sun}$, and $M_{\rm B}$ = 0.37 $M_{\sun}$, $L_{\rm B}$ = 0.17 $L_{\sun}$, respectively, measured from the optical spectroscopic observations \citep{Hartigan03,Herczeg14}.
The power-law index of the surface density $p$ is fixed to be the canonical value of -1.
The inner cutoff radius of both the dust and gas disks is fixed to be 0.01 au.
The outer cutoff radius of the gas disks is adopted to be 260 au, which is taken from the outermost extent of the observed CO emission at a 3$\sigma$ level in VLSR = 4.9 km s$^{-1}$ (blueshifted emission) and 7.4 km s$^{-1}$ (redshifted emission).
From our investigation of the P-V diagrams, the systemic velocities of XZ Tau A and B are both fixed to be $v_{\rm sys} \sim$ 6.0 km s$^{-1}$ (see Figure \ref{figs_pv_obs}).
As described in Section \ref{subsec:12CO}, the effect of the missing fluxes prevents us from accurately determining the systemic velocity of each CSDs.
Within this limitation, the common systemic velocity of 6.0 km s$^{-1}$ for both XZ Tau A and B yields disk models consistent with the observed image cube.
The position angles of the disks around XZ Tau A and B are set to be $\theta_A$ = -5.4$\degr$ and $\theta_B$ = 130.0$\degr$, respectively.
These position angles are derived from our manual inspection of the P-V diagrams made at a step of the position angle of 5$\degr$, and this 5$\degr$ step can be regarded as the error of the position angles (Section \ref{subsec:12CO}).
The turbulent velocity in the disks is set to be 0.4 km s$^{-1}$, adopted from our previous study of L1551 NE \citep{takakuwa12}.
The canonical abundance of $^{12}$CO, $X(^{12}{\rm CO})$ = 9.5$\times$10$^{-5}$, the gas-to-dust mass ratio of 100, and the H$_2$ ortho-to-para ratio of 3.0, are adopted.
These fixed parameters are summarized in the upper rows of Table \ref{table:model_parameters}.

We then adopt $\lambda$, $\Sigma_0$, disk inclination angle $i$, and the outer cutoff radius of the dust disks ($\equiv r_d$)
as free parameters.
These four parameters along with the fixed parameters determine the 1.3-mm dust-continuum flux densities.
We varied these four parameters and searched for the parameters which reproduce the observed 1.3-mm dust-continuum flux densities.
The parameter search was made for XZ-Tau A and B separately.
For each parameter set, we performed radiative transfer calculations using the RADMC3d code \citep{radmc3d}.
The temperature distribution is first calculated by RADMC3d with the thermal Monte Carlo method. 
Then with the assumed density distribution and the calculated temperature distribution, the radiative transfer calculations were conducted on the assumption of the local thermodynamic equilibrium (LTE) condition, and the model flux densities of the 1.3-mm dust continuum emission are calculated.
Since the observed 1.3-mm continuum images are point sources, we can directly compare the model flux densities calculated with RADMC3d to the observed flux densities.
The inclination angles are varied in the range from 40$\degr$ to 85$\degr$, $\lambda$ from 1.0 to 1.7, $r_{{\rm d}}$ from 5.0 au to 8.0 au,
and $\Sigma_{{\rm dust},0}$ from 70 g cm$^{-2}$ to 100 g cm$^{-2}$.
We found that $\lambda$ need to be 1.3 and 1.5 in XZ Tau A and B, respectively, to match the model 1.3-mm continuum flux densities with the observed flux densities.
The best-fit parameters are summarized in the bottom-row of Table \ref{table:model_parameters}.
Since the absolute flux uncertainty of ALMA observations at Band 6 is 10$\%$, the ranges of the parameters which give the continuum flux densities of $\pm$10$\%$ from the observed 1.3-mm continuum flux densities are defined as the errors of the relevant parameters.

Once all the values of the model parameters are obtained as above, the model CO velocity channel maps are created.
The derived $\Sigma_0$ value of dusts can be converted to the $^{12}$CO, ortho-H$_2$, and para-H$_2$ surface densities at $r_0$ to be $\Sigma^{^{12}{\rm CO}}_{0}$ = 2.2$\times$10$^{23}$ cm$^{-2}$, $\Sigma^{\rm ortho-H_2}_{0}$ = 1.7$\times$10$^{27}$ cm$^{-2}$, and $\Sigma^{\rm para-H_2}_{0}$ = 5.7$\times$10$^{26}$ cm$^{-2}$, respectively.
The distributions of the dust density, $^{12}$CO number density, and the calculated temperature are shown in Figure \ref{fig:model_parameter}.
With this model setting, we performed a final RADMC3d calculation with the large velocity gradient (LVG) mode.
The individual model images of XZ Tau A and B were co-added to make the combined image.
Then, to incorporate the effect of the interferometric filtering effect, the ALMA observing simulation using the CASA task simobserve was conducted.
In simobserve, the same antenna configuration, hour angle coverage, bandwidth and frequency resolution, and integration time as those of the real ALMA observation are adopted.
The model visibility made by simobserve was CLEANed and deconvolved, and the final $^{12}$CO model velocity channel maps were created.
For a direct comparison between the observed and model images, in the models the noise is not included.
As it is not straightforward to properly model the outflow component, we do not include the outflow component in the model either.

Figure \ref{figs_12CO_model} shows the model velocity channel maps of the $^{12}$CO (2--1) emission.
In the high blueshifted velocities of 0.4--1.1 km s$^{-1}$, an emission component located between the binary is seen, and the component shifts toward the northeast from 1.7 km s$^{-1}$ to 3.0 km s$^{-1}$.
In the lower blueshifted velocities of 3.6--4.9 km s$^{-1}$, the $^{12}$CO emission peaks to the northeast of the binary are embedded in the extended emission.
On the other hand, in the high redshifted velocities of 9.3--11.2 km s$^{-1}$, two peaks located to the south of XZ Tau A and to the northwest of XZ Tau B are seen.
These features in the model velocity channel maps are consistent with those in the real, observed velocity channel maps (Figure \ref{figs_12CO_obs}). 
The model velocity channel maps exhibit extended emission components, originated from the butterfly components of the two disks, in the lower-velocity range of 5.5--8.0 km s$^{-1}$. 
In the real observed velocity channel maps, on the other hand, the $^{12}$CO emission in these velocities are significantly suppressed.
This is likely due to the effect of the missing fluxes caused by the presence of the extended cloud components, which are not included in the model.
We also note that the apparent circular features as seen in Figure \ref{figs_12CO_model} are originated from the outer cutoff in the disk models.

Figure \ref{figs_pv_model} compares the observed (left panel) and model (right) P-V diagrams of the $^{12}$CO (2--1) emission along the disk major axes.
Along the north-south direction passing through XZ Tau A, the major axis of the XZ Tau A disk, the observed Keplerian rotation signature is reproduced with our model (Figure \ref{figs_pv_model}a and d).
Along the northwest to southeast direction passing through XZ Tau B or the major axis of the XZ Tau B disk, the observed Keplerian rotation signature of the XZ Tau B disk is also reproduced in the model (Figure \ref{figs_pv_model}b and e). 
Furthermore, in the redshifted side the observed two emission protrusions toward the high velocities are seen in the model. 
The northern and southern redshifted protrusions correspond to the Keplerian rotation components of the XZ Tau B disk and XZ Tau A disk, respectively.

The above comparison between the observed and model velocity channel maps and P-V diagrams imply a misaligned configuration between the two Keplerian CSDs.
The differences in the position and inclination angles are $\sim$ 135$\degr$ and $\sim$ 4$\degr$, respectively.
We also note that the present modeling implies the same systemic velocities of XZ Tau A and B.

\subsection{Physical mechanism to Produce Misaligned CSDs}

The above modeling demonstrates that the CSDs in the binary system XZ Tau are misaligned with each other.
Recent high-resolution observations of young multiple systems have also found such a misaligned configuration of CSDs. ALMA DSHARP observations of the triple Class II system AS 205 found that
the circumbinary disk in AS 205 S and the circumstellar disk in AS 205 N, with the projected separation of $\sim$165 au, are misaligned, and that the difference of the inclination angles is $\sim$ 45$\degr$ \citep{Kurtovic18}.
The same DSHARP observations identified that the three CSDs as seen in the 1.3-mm dust-continuum emission around HL Lup A, B, and C, are all misaligned.
Furthermore, the rotational directions of the CSDs around HL Lup A and B, with the projected separation of $\sim$ 20 au, are almost opposite as seen in the $^{12}$CO (2--1) emission. 
Another ALMA observations of the binary system HK Tau revealed that the rotational vectors of the two CSDs, with the projected separation of 386 au, are misaligned by 60--68$\degr$ \citep{Jensen14}.
Such misaligned configurations are also found toward younger, Class I binary systems.
ALMA observations of the Class I binary IRS 43 have found that the two CSDs are significantly misaligned ($>$60$\degr$) \citep{Brinch16}. 
Toward L1551 NE, a Class I binary located to the south of XZ Tau, the two CSDs and the common CBD are all misaligned with each other \citep{takakuwa17}.

These results, unveiled by the latest high-resolution observations, suggest that misaligned binary/multiple systems are rather ubiquitous.
The presence of such systems in both Class I and II sources implies that misalignment of the CSDs should be developed in protostellar stages.
If a binary system is formed from a fragmentation of a common rotating disk, precursor of the CBD, the rotational vector of the two CSDs should be aligned with that of the natal CBD \citep{Nakamura03,Machida08}. 
On the other hand, recent observational and theoretical studies suggest that cloud cores are turbulent without significant systematic angular momentum vectors, and that the turbulence produces local angular momenta to form disks around protostars \citep{Takaishi20}.
If binary protostellar systems are formed through turbulent fragmentations in such dense cores, followed by the subsequent formation of the CBD, the misalignment between the CSDs as well as that between the CSDs and the CBD are naturally explained \citep{Padoan02,Bate10,Offner10}.
Toward a single Class I protostar L1489 IRS, \cite{Sai20} have found that there is a gap in the Keplerian disk at a radius of $r \sim$ 200 au as seen in the C$^{18}$O (2--1) emission, and that inside and outside the gap the Keplerian disk is misaligned by $\sim$ 15$\degr$.
Formation of such a warped disk around a single star could also be reproduced if the vector of the local angular momentum of the infalling material changes, as demonstrated by their theoretical model.

Another possible mechanism to produce misalignments of CSDs in binary systems is Kozai-Lidov (KL) mechanism \citep{Kozai62,Lidov62}.
The KL mechanism produces periodic, mutual exchange the eccentricity and inclination of a CSD in one component of the binary systems, $i.e.$, KL oscillation.
For the KL oscillation to occur, however, the initial inclination of a particle orbit around one component of the binary with respect to the binary orbital plane ($\equiv i_0$) should be $\cos^2(i_0) < \frac{3}{5}$ or $39\degr < i_0 < 141\degr$ \citep{Martin14,Fu15}.
Theoretical simulations of gas disks around one component of a binary show a somewhat larger critical inclination ($> 45\degr$) \citep{Fu15}.
Thus, it is unlikely that initially aligned CSDs in a binary system turn into misaligned CSDs via the KL mechanism.
On the other hand, the derived disk inclination angles of the CSDs in XZ Tau from the plane of the sky is $> 60\degr$ (Table \ref{table:model_parameters}), and the orbital plane of the binary is likely the plane of the sky as described in the next subsection.
The inclinations of the CSDs are thus above the critical angle for the KL oscillation to occur.
The timescale of the KL oscillation ($\equiv \tau_{\rm KL}$) is described as
\begin{equation}
    \tau_{\rm KL} \approx \frac{M_1+M_2}{M_2} \frac{P_{\rm b}^2}{P_{\rm p}} \left( 1-e_{\rm b}^2 \right)^{\frac{3}{2}},
\end{equation}
where $M_1$ and $M_2$ are the masses of the primary and secondary stars, $P_{\rm b}$ is the orbital period of the binary, $P_{\rm p}$ that of the CSD, and $e_{\rm b}$ is the eccentricity of the binary orbit \citep[$e.g.$][]{Kiseleva98,Martin14}.
With the binary orbital parameters derived in the next subsection,
the timescale of the KL oscillation of XZ Tau at the radius of the dust disk ($r_d$ = 7 au) is calculated to be $\sim$ 376 yr, 2.4 times the orbital period of the binary system XZ Tau.
It is possible that the XZ Tau binary system experiences the KL oscillation, after the misaligned system is produced via the turbulent fragmentation.

If misaligned CSDs in binary systems are formed in the protostellar stages, they should be present until the Class II stages.
Theoretical work by \cite{Bate00B} showed that tidal shearing and viscosities substantially reduce the alignment time scale of the CSDs in binary systems.
In such a case, external mechanisms, such as gravitational interaction of a passing object or infall of materials with different angular momenta from the CBDs \citep{Smallwood21}, are required to reproduce the observed misaligned T-Tauri binary systems.
On the other hand, 3D hydrodynamic simulations by \cite{Fragner10} have shown that for thick disks with $h$ = 0.05 and a low viscosity $\alpha$,
the alignment timescales in binary systems are long enough.
Thus, the observed misalignments of the disks in the T-Tauri binaries could survive from their formation in the protostellar phase until the disk lifetime.

\subsection{Orbital Solution of the Misaligned Binary} \label{subsec:orbit}

\begin{figure*}
\centering
\includegraphics[width=80mm, angle=0]{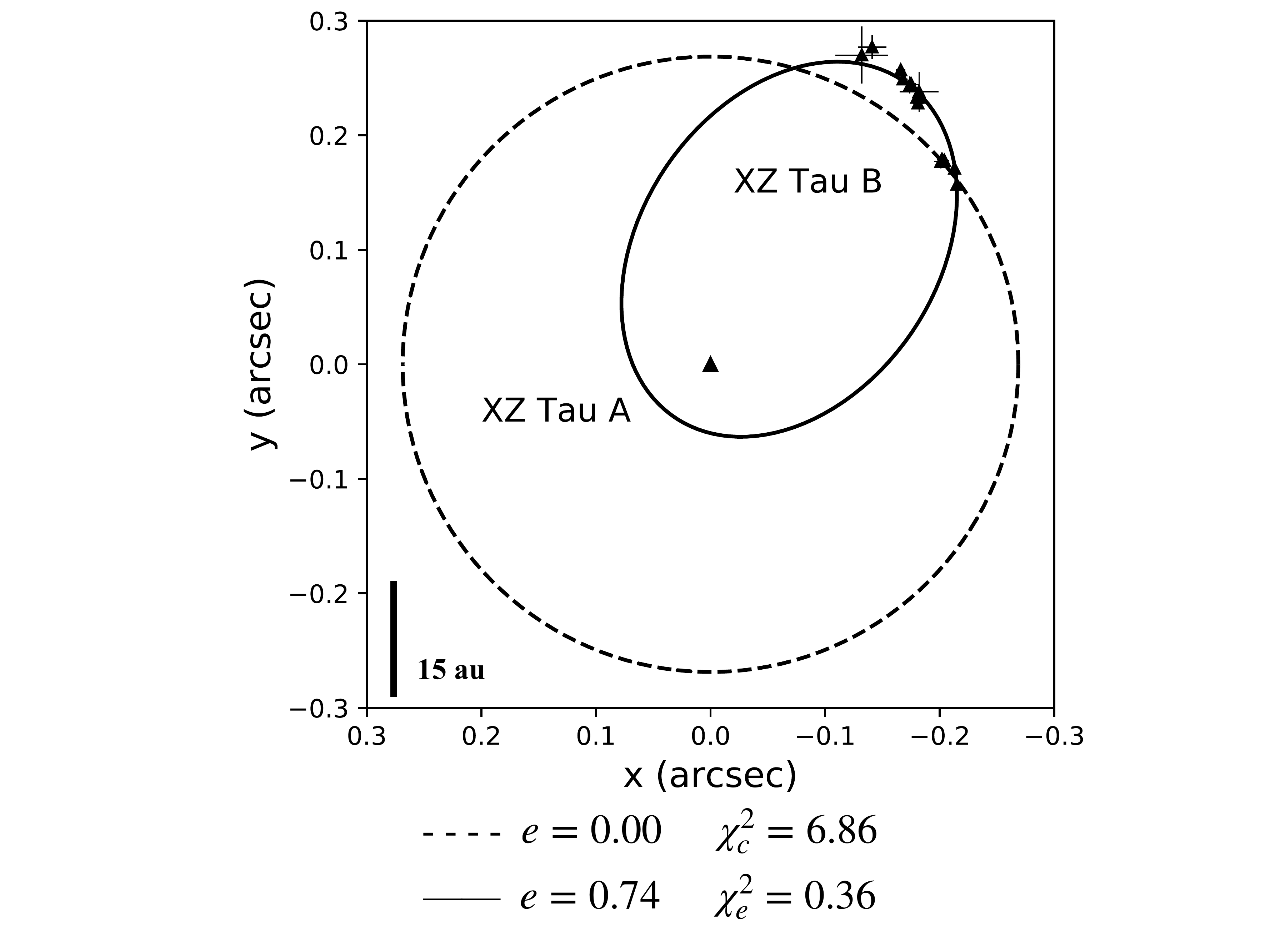}
\caption{Estimated orbits of XZ Tau B with respect to XZ Tau A. 
The origin of the coordinate system is in the XZ Tau A position (central triangle).
Triangles with error bars show the positions and the errors of XZ Tau B.
Dashed and solid curves show the circular and elliptical orbits estimated from the least-square fitting.
The orbital eccentricity $e$ and the fitting $\chi^2$ values are shown below.
}
\label{figs_Orbit}
\end{figure*}

\begin{figure}
\centering
\includegraphics[width=80mm, angle=0]{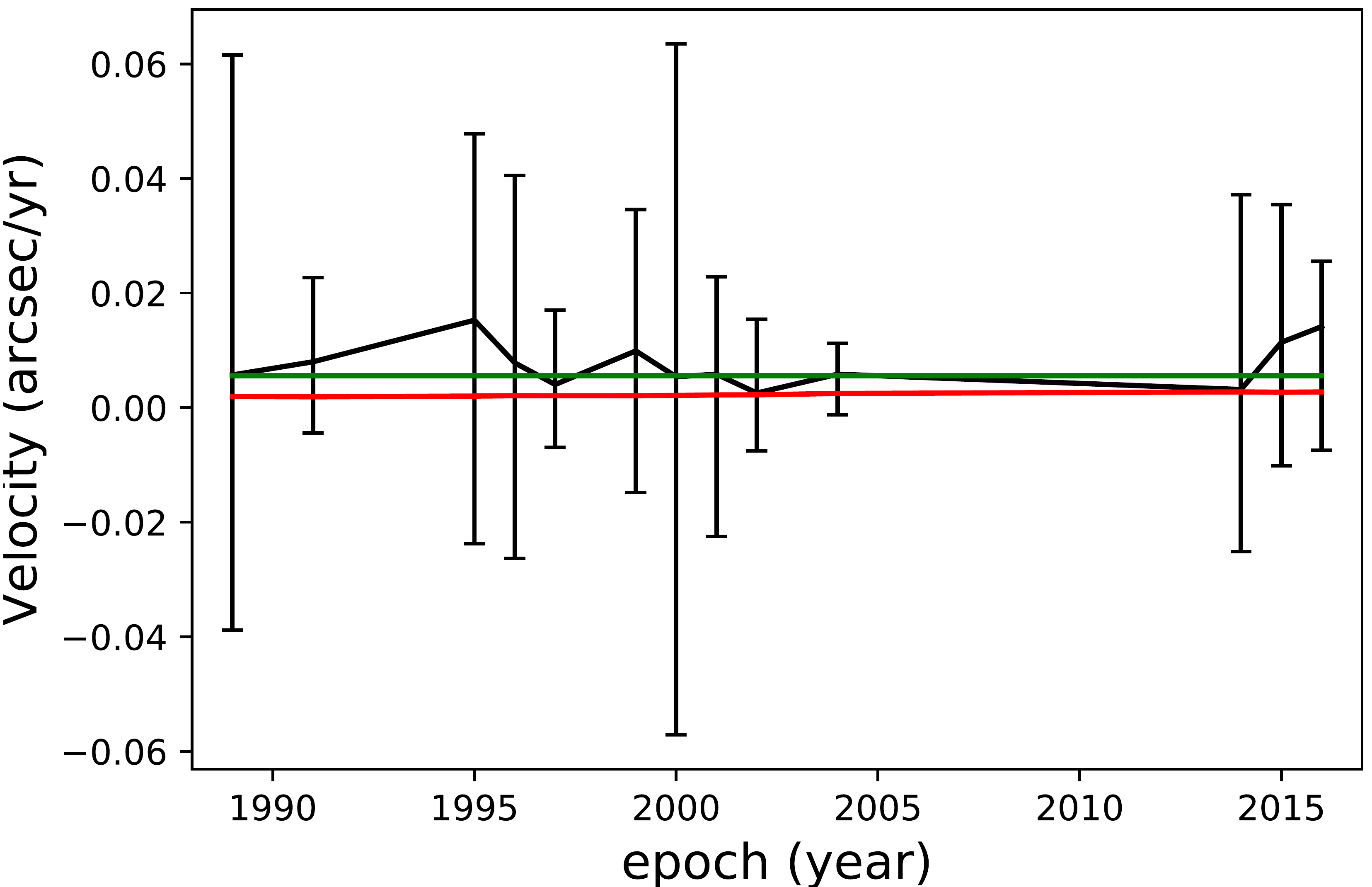}
\caption{Measured transverse velocities of XZ Tau B with respect to XZ Tau A.
The black line with the error bars shows the velocities estimated from the NIR and ALMA observations.
The green and red lines show the velocities estimated from the best-fit circular and elliptical orbits, respectively, shown in Figure \ref{figs_Orbit}.
}
\label{figs_velocity}
\end{figure}

We have unveiled clear positional shifts of XZ Tau B with respect to XZ Tau A from the 1.3-mm dust-continuum images observed in 2015--2017.
\cite{Dodin16} have also found such positional shifts of XZ Tau B with respect to XZ Tau A from the NIR observations in 1989--2014, and discussed possible orbital solutions of the binary.
The ALMA positional shifts are in good agreement with their results, and we here discuss the possible orbital solution using both the previous NIR data and the present ALMA data.

From our analyses and modeling of the P-V diagrams (see Figure \ref{figs_pv_obs} and \ref{figs_pv_model}), the systemic velocities of both CSDs in XZ Tau A and B are found to be almost identical to be $V_{\rm LSR}$ = 6.0 km s$^{-1}$.
Thus, it is natural to assume that the orbital plane is on the sky ($i.e.$, $i$ = 0.0$\degr$), even though in rare occasions it is possible to have the same line-of-sight velocities of the binaries in the inclined orbital plane.
Based on the face-on assumption, the $\chi^2$ minimizations have been performed to derive the best-fit orbital solutions in two cases, circular and elliptical orbits.
In the case of the circular orbit the $\chi^2$ minimization was performed with the common radius as a fitting parameter, whereas in the case of the elliptical orbit three fitting parameters, semi-major axis ($\equiv a$), eccentricity ($e$), and the argument of pericenter ($\omega$), are adopted. 
The derived best-fit orbital solutions and the fitting $\chi^2$ values are shown in Figure \ref{figs_Orbit}.

In Figure \ref{figs_Orbit}, it is clear that the fitting result in the case of the circular orbit is worse than that of the elliptical orbit.
As the positional accuracy of the present ALMA data is much better than that of the NIR data, the orbital curve derived from the $\chi^2$ fitting in the case of the circular orbit traces the ALMA data selectively, and the NIR data points are systematically located outside the circle.
On the other hand, the elliptical orbit with $e$ = 0.74 traces both the NIR and ALMA data points equally well.
These results imply that the addition of the new ALMA data constrains the possible orbit better than the previous NIR data alone.
Furthermore, the ALMA data demonstrates that the elliptical orbit is better than the circular orbit.
The best-fit orbital parameters are $e$ = 0.742$^{+0.025}_{-0.034}$ at $i$ = 0.0$\degr$, $a$ = 0$\farcs$172$^{+0\farcs002}_{-0\farcs003}$ (25 au), and $\omega$ = -54.2$^{+2.0}_{-4.7} \degr$.
The orbital period is then $\sim$ 155 yr.

Figure \ref{figs_velocity} shows the measured transverse velocities of XZ Tau B with respect to XZ Tau A on the plane of the sky.
Here, the velocities at the individual epochs are defined from the average velocity using the closest two data before and after that epoch, and at the first and last epochs only the data next and before that epoch are used, respectively.
The error bars of the velocity are calculated from the propagation of the positional errors.
The derived velocities are on average $\sim$ 7.6$\times$10$^{-3}$ arcsec yr$^{-1}$, and a systematic trend of the velocity, required to derive the orbital solution based on the law of constant areal velocity, is not identified because of the large error bars.
The measured transverse velocities are approximately consistent with those of the anticipated orbital velocities
in the elliptical orbit (red line in Figure \ref{figs_velocity}) and those in the circular orbit (green line).

\begin{figure}
\centering
\includegraphics[width=80mm, angle=0]{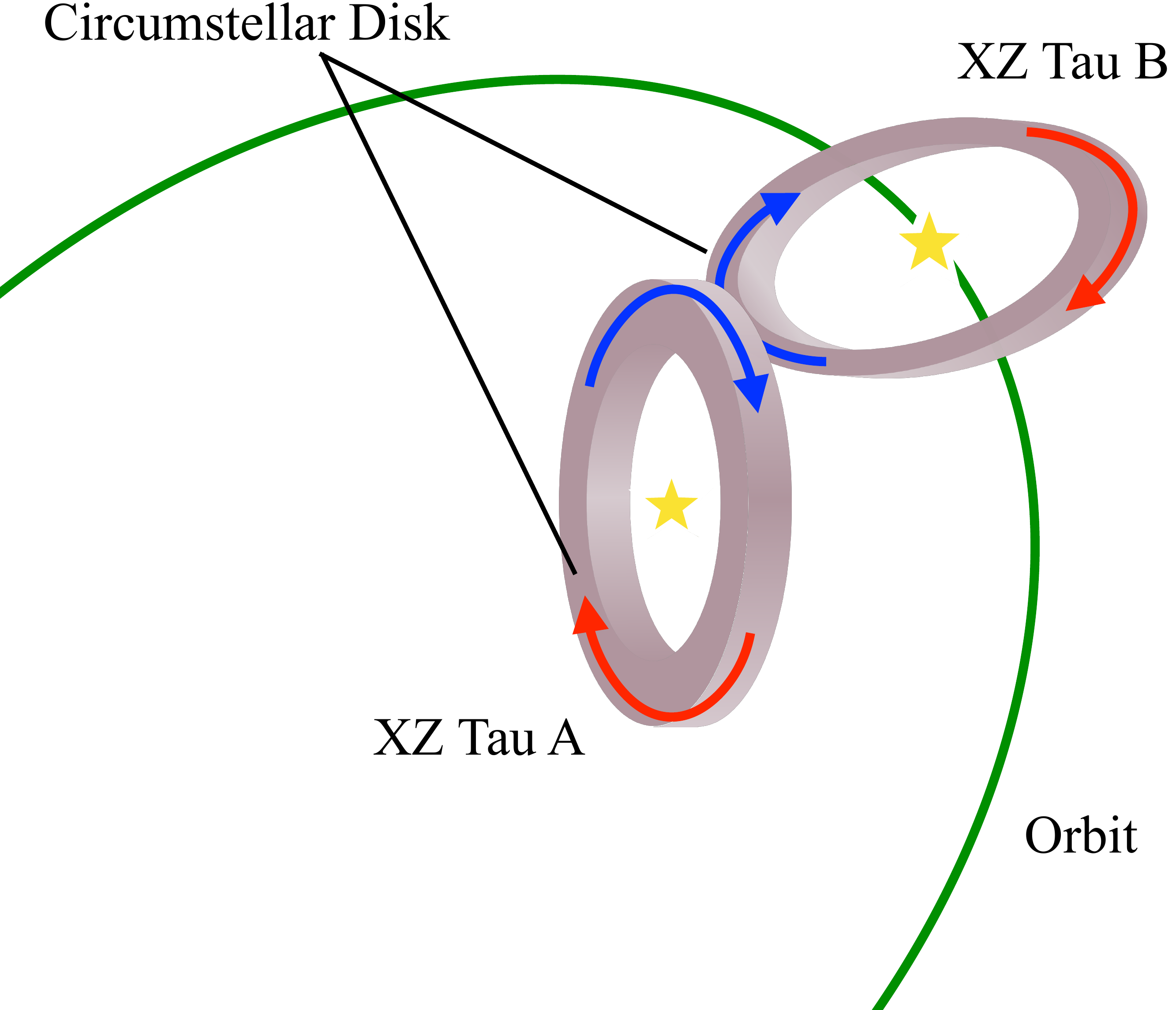}
\caption{Schematic picture of the binary system XZ Tau revealed by the present ALMA observations.}
\label{figs_image}
\end{figure}

The orbital solution, along with the misaligned two CSDs, suggests that the two CSDs and the orbital plane are all misaligned.
A schematic illustration of our results is shown in Figure \ref{figs_image}.
Whereas recent high-resolution observations have found misaligned CSDs in binary systems as described in the last subsection, or orbital motions of the binary systems \citep[$e.g.$][]{Lim16a,Lim16b}, our analyses have unveiled the misaligned CSDs, misaligned orbital plane, and the elliptical orbit, all at the same time.
This result should support the fragmentation of turbulent dense cores as the formation mechanism of the binary \citep{Bate00A,Offner10}.
Furthermore, SPH simulations by \cite{Bate00A} have shown that the eccentric binary orbit can be formed through the turbulent fragmentation of dense cores.

Our multi-epoch analyses have proven that we can conduct time domain science with the wealth of the ALMA archival data, "ALMA movie".
For the next $\sim$ 20 years, the XZ Tau A and B are expected to be aligned almost east-west, and we should be able to confirm the eccentric orbit.
Other important time-domain sciences feasible with the ALMA archival data are monitoring of the proper motions of molecular jets driven from protostars \citep{Girart01,Yoshida21}, and monitoring of protostellar luminosities and accretion bursts \citep{Johnstone18,Lee20}.
These time domain science should become more important as more and more observational data have been accumulated.

\section{Summary} \label{sec:summary}

We have analyzed multi-epoch (2015--2017) ALMA archival data of the Class II binary system XZ Tau in the 1.3-mm, 2.1-mm, and 3.5-mm dust-continuum emission and the $^{12}$CO ($J$=2--1) line.
Our main results are summarized below.

\begin{itemize}
    \item [1.]
    The dust-continuum emission traces two circumstellar disks (CSDs) around the individual binary stars (XZ Tau A and B) but the circumbinary disk (CBD), which is seen in the neighboring Class I binary systems of L1551 IRS 5 and NE.
    The dust emission originated from the CSDs are unresolved even in the highest-resolution ($\sim$0$\farcs$03) 1.3-mm image, and the upper limit of the dust disk sizes is $\lesssim$15 au.
    The 1.3-mm flux densities originated from the CSDs in XZ Tau A and B are $\sim$ 5.8 mJy and 7.3 mJy, which gives the disk masses of $\sim$ (0.77--3.47)$\times$10$^{-3}$ $M_{\odot}$ and (0.96--4.36)$\times$10$^{-3}$ $M_{\odot}$, respectively, on the assumption of $\kappa_{\rm 1.3mm}$ = 0.019 cm$^2$ g$^{-1}$ for the dust temperature $T_{\rm d}$ = 10--30 K.
    
    \item [2.]
    The $^{12}$CO ($J$=2--1) data taken on 6 Oct. 2016 reveal the detailed spatial and velocity structures of the molecular gas around the binary system.
    In the highest blueshifted velocity ($V_{\rm LSR}$ = -0.7 - 0.3 km s$^{-1}$), two emission components are seen to the north of XZ Tau A and the southeast of XZ Tau B.
    The redshifted (11.0--12.7 km s$^{-1}$) counterparts to these high-velocity blueshifted emission are seen to the south of XZ Tau A and the northwest of XZ Tau B.
    In the intermediate velocity ranges, the blueshifted (-0.1--3.2 km s$^{-1}$) emission is shifted to northeast of the binary and the redshifted emission (8.1--10.8 km s$^{-1}$) southwest.
    The Position - Velocity diagram along the north-south direction passing through the highest blueshifted emission, XZ Tau A, and the highest redshifted emission shows the Keplerian rotation signature.
    The P-V diagram along the northwest to southeast passing through XZ Tau B and the surrounding highest velocity emission reveals presence of another Keplerian rotation signature.
    Comparison of these results with our RADMC3d model of Keplerian disks shows that those components around XZ tau A and B trace the CSDs of the individual stars, and the rotational directions of the CSDs of XZ Tau A and B are misaligned with each other.
    The differences of the position and inclination angles are $\sim$ 135$\degr$ and $\sim$ 4$\degr$, respectively.
    In the intermediate and lower velocity ranges, on the other hand, another blue- (-0.1--5.3 km s$^{-1}$) and redshifted (7.0--8.8 km s$^{-1}$) emission appear to the southwest and the northeast of the binary, respectively.
    These $^{12}$CO emission likely traces the associated molecular outflow.
    
    \item [3.]
    From the multi-epoch ALMA archival data, a systematic positional shift of the CSD in XZ Tau B with respect to that in XZ Tau A is unveiled.
    The relative position is moving toward southwest, and the detected shift from 2015 to 2017 is $\Delta$R.A. $\sim$ -1.6 au and $\Delta$Dec $\sim$ -3.0 au.
    The sense of the positional shift detected with ALMA is consistent with that of the previous NIR observations from 1989 to 2014.
    Due to the large errors it is not possible to derive a clear trend of the change of the moving velocity as a function of time.
    
    \item [4.]
    As the systemic velocities of the CSDs around XZ Tau A and B are similar and both $\sim$ 6.0 km s$^{-1}$, the orbital plane is likely parallel to the plane of the sky ($i$ = 0.0$\degr$).
    We then estimated possible orbital solutions of the binary on the plane of the sky with the $\chi^2$ fitting of the circle / ellipse to the observed positions.
    We found that an eccentric orbit ($e \sim$ 0.7) is preferable to reproduce the observed trajectory, and the circular orbit ($e =$ 0.0) is likely excluded, as the $\chi^2$ value in the case of the circular orbit ($\chi_c^2$ = 6.86) is an order of magnitude worse than that of the elliptical orbit ($\chi_e^2$ = 0.36).
    The best-fit orbital parameters are
    $e$ = 0.742$^{+0.025}_{-0.034}$ at $i$ = 0.0$\degr$, $a$ = 0$\farcs$172$^{+0\farcs002}_{-0\farcs003}$ (25 au), and $\omega$ = -54.2$^{+2.0}_{-4.7} \degr$.
    The orbital period is derived to be $\sim$ 155 yr.
    The observed transverse velocities are also consistent with those anticipated from the elliptical orbit on the plane of the sky.
    Subsequent monitoring observations of XZ Tau should strengthen the presence of the eccentric orbit, and our analyses of the ALMA archival data demonstrate the scientific importance of time-dependent science with ALMA, "ALMA movie".

    \item [5.]
    Our results, as well as recent high-resolution observations of young binaries, suggest that misaligned binary / multiple systems are ubiquitous.
    These results imply that turbulent fragmentations in natal dense cores, not rotationally-driven fragmentation of a common disk, are the likely mechanism to form misaligned binary systems.
    Our new ALMA results of the binary system XZ Tau have found that not only the CSDs but also the orbital plane are misaligned with each other, and that the orbit is likely eccentric. These results further strengthen the turbulent fragmentation scenario of binary formation.
    
\end{itemize}

\acknowledgments
We are grateful to the anonymous referee for insightful comments and detailed reading of the manuscript.
We would like to thank all the ALMA staff supporting this work.
S.T., Y.T., and Y.S. are supported by JSPS KAKENHI grant Nos.
18K03703, 21H00048, 21H04495, and 
19K23463, JP20K04035, and 18K13581, 18H05437, respectively.
This paper makes use of the following ALMA data:
ADS/JAO.ALMA\#2013.1.00105.S, \#2016.1.00138.S, and \#2017.1.00388.S.
ALMA is a partnership of ESO (representing its member states), NSF (USA), and NINS (Japan), together with NRC (Canada), MOST and ASIAA (Taiwan), and KASI (Republic of Korea), in cooperation with the Republic of Chile. The Joint ALMA Observatory is operated by ESO, AUI/NRAO, and NAOJ.






\end{document}